\documentclass[11pt]{article}

\usepackage[margin=1in]{geometry}
\usepackage{amsmath,amssymb}
\usepackage{graphicx}
\usepackage{booktabs}
\usepackage{xcolor}
\usepackage{titlesec}
\usepackage[colorlinks=false,allcolors=blue]{hyperref}
\usepackage{lineno}
\usepackage{pdflscape}
\usepackage[
    style=ieee,
    backend=biber,
    sorting=none
]{biblatex}

\graphicspath{{figures/}}

\titleformat{\section}
  {\centering\normalsize\scshape}
  {}{0pt}{\MakeUppercase}

\newcommand{\runin}[1]{\medskip\noindent\textit{#1}\ }

\title{A nonlinear dynamics approach to assess amplitude-frequency coupling in non-invasive human electrogastrography recordings}

\author{%
\parbox[t]{0.45\textwidth}{%
    \centering
    \textbf{Akshita A. Rao}\\[0.3em]
    Dept. of Bioengineering\\
    Stanford University\\[0.3em]
    \href{mailto:akshitar@stanford.edu}{akshitar@stanford.edu}
}
\hfill
\parbox[t]{0.45\textwidth}{%
    \centering
    \textbf{Todd P. Coleman}\\[0.3em]
    Dept. of Bioengineering\\
    Stanford University\\[0.3em]
   \href{mailto:toddcol@stanford.edu}{toddcol@stanford.edu}
}
}

\date{}
\begin{document}

\maketitle

\begin{abstract}
\noindent
Conventional electrogastrography (EGG) analyses summarize gastric slow-wave activity using dominant frequency, power, or phase-based metrics, without explicitly modeling amplitude-dependent frequency dynamics. We developed a statistical nonlinear-dynamics framework that represents the gastric-band analytic EGG signal as a stochastic Stuart-Landau oscillator and estimate local effective phase-shear, $\beta_{\mathrm{eff}}$, a normalized measure of amplitude-frequency coupling. In overnight recordings from 60 healthy participants, effective phase-shear was consistently negative, generalized to held-out halves of each recording, and was abolished by amplitude surrogates that disrupted amplitude-phase alignment while preserving amplitude-series structure. We then applied the measure to fasted and fed recordings from 16 healthy participants and 24 participants with gastric dysfunction. Phase-shear magnitude differed between groups and classified dysfunction comparable to the strongest multielectrode phase-based traveling-wave measures. These findings establish amplitude-frequency coupling as a dynamical feature of gastric-band EGG that complements conventional spectral and spatial phase metrics, while motivating future multichannel observation models to determine its relationship to slow-wave propagation and gastric function.
\end{abstract}

\vspace{0.5em}
\noindent

\vspace{0.6em}
\noindent

\section{Introduction}

The gastric slow wave is an electrical rhythm generated by interstitial cells of Cajal (ICC) in the myenteric plexus, with a nominal frequency near 3 cycles per minute (cpm; $\sim$0.05\,Hz) in humans \cite{sandersInterstitialCellsCajal2006}. It coordinates the timing and propagation of gastric electrical activation and thereby constrains the spatiotemporal organization of gastric contractions. The electrogastrogram (EGG) noninvasively records gastric myoelectrical activity from cutaneous abdominal electrodes \cite{wolpertElectrogastrographyPsychophysiologicalResearch2020c}. Studies have linked conventional EGG abnormalities, including altered dominant-frequency distributions, frequency instability, and abnormal postprandial power responses, to delayed gastric emptying and gastroduodenal disorders \cite{chenAbnormalGastricMyoelectrical1996, riezzoElectrogastrographyAdultsChildren2013, palmezanoElectrogastrographyAdultGastroparesis2025, gharibansSpatialPatternsHighResolution2019c, geldofMyoelectricalActivityStomach1989, ogradyAbnormalInitiationConduction2012a}. However, the microvolt level cutaneous signal is susceptible to respiratory and cardiac contamination, motion, postural changes, and electrode contact. Further, the waveform from an abdominal electrode is a spatially aggregated projection of  nearby ICC cells, rather than the direct output of a single oscillator \cite{wolpertElectrogastrographyPsychophysiologicalResearch2020c, ArtifactRejectionMethodology}. Despite these confounds, the EGG is the most widely used noninvasive marker of gastric myoelectric activity \cite{ogradyOriginPropagationHuman2010}.

Mathematical approaches to gastric-slow wave recordings can be broadly grouped into mechanistic models of slow-wave generation and propagation, including coupled-oscillator \cite{perleyGraphicalModelsEfficient2025, hwangKuramotovonMisesTime2026} and biophysically based tissue models \cite{linkensStabilityEntrainmentConditions1977, sellier-pronoDefectsParcellationRenormalized2025, Peng2018}, and statistical signal-analysis approaches that summarize cutaneous EGG using dominant frequency, spectral power, frequency instability, or phase-based measures \cite{wolpertElectrogastrographyPsychophysiologicalResearch2020c,gharibans2016high}. Mechanistic models have provided important insight into ICC-network behavior and spatial propagation, but are commonly parameterized from cellular, serosal, or simulated data rather than fitted to cutaneous EGG recordings according to a statistical model. Conversely, conventional  analyses of human EGG recordings characterize the signal in finite windows through time-averaged or windowed spectral properties, such as dominant frequency, dominant power, the percentage of time spent in frequency ranges, and frequency-instability measures. These summaries need not assume that the biological oscillator is strictly harmonic; however, they do not explicitly model a dynamic relationship between the instantaneous amplitude of the processed EGG and its local phase velocity.

Neither a fixed-frequency nor a static-amplitude description fully captures continuous EGG data. Dominant frequency is conventionally summarized over successive analysis windows, and its within-record variation has long been quantified using the dominant-frequency instability coefficient, defined as the standard deviation of dominant frequency divided by its mean
\cite{kochEffectLiquidSolid2023, parkman2003}. This metric has been reported to differ between fasting and postprandial recordings and to be elevated in gastric dysrhythmia and gastroparesis cohorts \cite{parkman2003}. A single global rate cannot represent local frequency drift over long recordings; whether a locally varying baseline improves prediction is therefore an empirical question we aim to address below. Furthermore, recent work has demonstrated gastric-band power fluctuations occurring every 2-3 minutes that covaries with cortical sleep dynamics \cite{raoSimultaneousStomachbrainElectrophysiology2025a}, with additional studies supporting similar amplitude changes over time \cite{smoutWhatMeasuredElectrogastrography1980}. These observations suggest treating gastric-band amplitude as a potentially time-varying feature rather than only a scalar summary may be useful. However, because preprocessing can shape the structure of a filtered analytic envelope, amplitude dynamics require interpretation with matched controls. A phase-only description nevertheless does not explicitly quantify whether local amplitude fluctuations covary with local phase velocity. This motivates treating the processed EGG as an effective nonlinear oscillator in which amplitude and frequency may be coupled state variables rather than static summaries.

Mechanistic models have provided important insight into the generation and propagation of gastric slow waves, but their parameters are often chosen to reproduce selected simulated or invasive observations rather than estimated from cutaneous recordings with an explicit statistical model. This limits formal assessment of parameter uncertainty, sensitivity to preprocessing, and out-of-sample predictive performance. A useful precedent in another physiological oscillator has been statistical circadian models, which combined a nonlinear oscillator with explicit models of observation-level physiological and protocol-related contributions \cite{StatisticalModelHuman}. For cutaneous EGG, an analogous framework can treat the processed gastric-band signal as a nonlinear oscillator, observing its dynamical properties over time, while maintaining a distinction between fitted signal-level parameters and direct cellular physiology.

The Stuart-Landau equation, the normal form of a supercritical Hopf bifurcation, is the canonical weakly nonlinear oscillator for systems near the onset of self-sustained oscillation \cite{strogatzNonlinearDynamicsChaos2018}, and in Cartesian form is
\begin{equation}
\label{eq:deter_SL}
\frac{dZ}{dt} = (\mu + \mathrm{i}\omega_0)\,Z - (1 + \mathrm{i}\beta)\,|Z|^2 Z ,
\end{equation}
where $Z(t)$ is a complex-valued effective state that within this manuscript  represents the processed gastric-band EGG signal. Within this deterministic normal form, $\omega_0$ is a baseline angular frequency, $\mu$ controls radial stability, and $\beta$ governs amplitude-dependent frequency shift. If $\mu \le 0$ the origin is stable and the system does not oscillate autonomously. If $\mu > 0$ the system settles onto a stable limit cycle of radius $a^\ast = \sqrt{\mu}$, and $\mu$ governs the rate at which the solution returns to that cycle after perturbation. The real and imaginary parts of the cubic term therefore play different roles: the real part sets how large the oscillation grows, and $\beta$ sets how much its frequency shifts as it does. If $\beta = 0$ and the amplitude is held at its limit-cycle value, the model reduces to a simple harmonic oscillator. As it is not precisely known how the ICC network and its autonomic inputs produce the observed cutaneous rhythm, the terms of the Stuart-Landau equation cannot be given anatomic interpretations beyond the definitions of $\omega_0$, $\mu$ and $\beta$. We therefore propose the equation as an analytically tractable system capable of representing the amplitude-modulated nonlinear oscillation characteristic of the EGG.

We present a statistical nonlinear-dynamics framework for analyzing the human gastric slow wave in cutaneous electrogastrography (EGG). The gastric-band analytic EGG signal is represented as a stochastic Stuart-Landau oscillator, with slow baseline-frequency variation accommodated by a local time-varying baseline rather than a single global intercept. We validate the resulting effective-shear parameter,
$\beta_{\mathrm{eff}}$, in simulations and long overnight recordings, where local amplitude-frequency coupling generalizes to held-out halves of each recording. We then apply this framework to controlled fasted and fed recordings from healthy participants and patients with gastric dysfunction. At the participant level, effective-shear magnitude distinguishes gastric-dysfunction from healthy recordings and achieved discrimination that was comparable to the strongest multielectrode phase-based traveling-wave measures. These findings establish amplitude-frequency coupling as a dynamical feature of gastric-band EGG that complements conventional spectral and spatial phase metrics, while motivating future multichannel observation models to determine its relationship to slow-wave propagation and gastric function.

\section{Preliminaries and notation}
\label{sec:prelim}
All stochastic processes are defined on a filtered probability space
$(\Omega,\mathcal{F},\{\mathcal{F}_t\}_{t\ge 0},\mathbb{P})$ satisfying the usual conditions, and every process below is adapted to $\{\mathcal{F}_t\}$. A standard real Wiener process $W = \{W_t\}_{t \ge 0}$ has $W_0 = 0$, independent increments, and $W_t - W_r \sim \mathcal{N}(0, t-r)$ for $r < t$. Two such processes $W_1, W_2$ are independent, and $W = W_1 + \mathrm{i}W_2$ denotes the resulting complex Wiener process. All stochastic differential equations are read in the It\^o sense \cite{ito1951stochastic,oksendal2003stochastic}. We use capital letters for random variables and stochastic processes ($Z$, $A$, $S$, $Y$, $R$, $W$) and lowercase letters for deterministic quantities, whether a model parameter, a fixed constant or an index ($a^\ast$, $c_3$, $c$, $m$, $t$, $i$, $n$). Bold upright symbols denote vectors and matrices ($\mathbf{X}$, $\mathbf{K}$, $\mathbf{Y}$, $\boldsymbol\eta$, $\mathbf{x}_i$); light symbols denote scalars. Greek letters are exempt from the case rule, and this includes model parameters ($\mu$, $\omega_0$, $\beta$), errors ($\varepsilon$) and the phase ($\theta$) keep their conventional forms, and the phase $\theta(t)$ is random despite being lowercase. A small set of deterministic quantities retains conventional uppercase notation, such as the diffusion coefficient $D$, the record length $T$, the sample count $N$, and the reported statistics $R^2_{\mathrm{bin}}$, $\mathrm{SE}$ and $\mathrm{MAD}$. Finally, the regressors in Section~\ref{sec:methods} are functions of the observed data and are therefore random, but every regression below is understood conditionally on them; they are accordingly written as fixed design quantities ($\mathbf{X}$, $\mathbf{x}_i$, $A_i$ inside a design row), and the standardized shear $z_\beta$ retains its conventional lowercase as a test statistic.

\section{Methods and model formulation}
\label{sec:methods}
\subsection{Data cohorts and preprocessing}
Two EGG data cohorts were used for our model parameter estimation and application. An overnight sleep cohort was used for model development and served as a validation cohort due to the long duration of recordings. A prandial (pre- and post-meal) cohort was used to test the application of the model parameters to assess its utility in detecting changes in gastric slow wave dynamics. Additionally, we use ground-truth simulations to establish which parameters ($\omega_0$, $ \mu$, or $\beta$) can be confidently captured from our proposed methods. 

\runin{Overnight sleep cohort.} Four-channel EGG was recorded in 60 healthy participants during approximately 8-10\,hours of overnight sleep across 3 nights with concurrent polysomnography all sampled at 500 Hz \cite{schochreport2023}. One session is an undisturbed baseline night and the other two involve controlled awakenings based on the experimental protocol. Thus, this dataset yielded 144 usable EGG recordings that underwent semi-automatic artifact rejection \cite{raoSimultaneousStomachbrainElectrophysiology2025a}. Because the parameters of interest are properties of a single recording rather than of the awakening manipulation, the three sessions are pooled for model development and the baseline-versus-awakening comparison is reported as a null (Section~\ref{sec:results}).

\runin{Prandial cohort.} Multi-channel EGG was recorded during a controlled prandial protocol that consisted of a dry-fasting state, consumption of a meal, and a fed state. Data acquisition methods ranged from a high-spatial resolution 25-channel (sampled at 250 Hz) or 8-channel (sampled at 250\,Hz) EGG system \cite{gharibansSpatialPatternsHighResolution2019c} to a 16-channel (sampled at 125\,Hz) ambulatory EGG patch \cite{razzakFunctionalizedAdhesiveThin2025}. The analysis comprises 50 recordings from 40 participants: 26 healthy recordings and 24 patient recordings, the latter from patients with gastric dysfunction carrying a diagnosis of functional dyspepsia ($n = 7$) or gastroparesis ($n = 17$). The healthy set pools three sources and, unlike the patient set, contains repeated sessions from the same individuals: 7 single-session 25-channel controls, 11 8-channel recordings from 4 individuals, and 8 ambulatory recordings from 5 individuals, giving 26 recordings from 16 healthy participants. Every reported statistic collapses a participant's repeated sessions to one value before any summary or test, so the healthy arm contributes 16 units of inference rather than 26 (see the statistical methods below).

\runin{Preprocessing of the analytic signal.} The same pipeline was applied to both cohorts so that parameters are comparable across them. First, extragastric transients and baseline drift were suppressed by
Wiener-residual artifact rejection: a Wiener filter was applied in a 20-second sliding window and its output subtracted from the signal, retaining the residual \cite{ArtifactRejectionMethodology}. The filter noise floor was set to each channel's own variance rather than to a fixed value, making the operation invariant to global amplitude scaling. Epochs annotated as artifactual were replaced with missing values rather than interpolated. Next, the signal was band-pass filtered between 0.02 and 0.10\,Hz (1.2-6\,cpm) using a fourth-order zero-phase Butterworth filter applied in second-order-section form. This range is wider than the narrow, peak-centered filters often used for time-resolved EGG analysis \cite{wolpertElectrogastrographyPsychophysiologicalResearch2020c}. We used the wider band to avoid imposing a narrow spectral constraint on the amplitude and phase fluctuations subsequently analyzed by the dynamical model, while retaining separation from the principal respiratory and cardiac frequency ranges.

The analytic signal $Z(t) = A(t)e^{i\theta(t)}$ was obtained by Hilbert transform, giving the instantaneous amplitude $A(t)$ and phase $\theta(t)$. The analytic signal was decimated to a common working rate of 2\,Hz by polyphase resampling; ground-truth simulation confirms that decimation to this rate does not bias any of the reported parameters (Section~\ref{sec:results}). Then, because artifact masking fragments the records, each contiguous artifact-free run of at least 120\,s was band-passed, Hilbert-transformed and decimated separately, and the resulting regression rows pooled, so that no finite difference is ever taken across a gap. Finally, a single EGG channel was analyzed per recording, selected as the channel with maximal power in the narrow gastric band 0.03-0.07\,Hz (pre-meal power for the prandial cohort, so that channel selection is blind to the fed state). All amplitude and phase quantities are computed within that same channel; because the gastric slow wave propagates, one channel's amplitude rhythm is not assumed to describe another channel's carrier.

\subsection{Model formulation}
\runin{The electrogastrogram model.} We assume that for a given cohort, EGG data $Y_{t_1},\dots,Y_{t_N}$ are collected on $[0,T]$ with $t_n = n\Delta t$. Each measurement may be expressed as
\begin{equation}
\label{eq:decomp}
Y_{t_n} \;=\; S_{t_n} \;+\; m\;+\; V_{t_n},
\end{equation}
where $S_{t_n}$ is the gastric-band oscillation, $m$ is the mean of the signal, and $V_{t_n}$ collects the extragastric contributions such as baseline drift, respiration, cardiac activity, motion and electrode-contact transients. The mean $m$ and the slow component of $V_{t_n}$ are attenuated by Wiener-residual rejection, which suppresses the great majority of power below 0.01\,Hz. The remaining out-of-band contributions to $V_{t_n}$, namely, respiration near 0.2-0.3\,Hz and cardiac activity near 1\,Hz, are removed by a band-pass filter. Since motion and electrode-contact artifacts are not always mitigated by filtering, artifacts were manually annotated and masked. Thus, what gets fed to our proposed model fit is $S_{t_n}$.

\runin{The stochastic Stuart-Landau representation.} Equation~\eqref{eq:deter_SL} defines a deterministic system. To capture the inherent fluctuations of real-world biological systems, we transition to the stochastic Stuart-Landau form (Eq.~\eqref{eq:stoch_SL}) \cite{perez-cerveraUniversalDescriptionStochastic2023}. The gastric signal \(S_{t_{n}}\) in \eqref{eq:decomp} is modeled as the real part of the complex state variable: 
\begin{equation}
\label{eq:SintermsofZ}
S_{t_n} = \mathrm{Re}\left\{Z(t_n)\right\}
\end{equation}

where Z(t) is governed by the It\^o stochastic differential equation \cite{ito1951stochastic,oksendal2003stochastic}:
\begin{equation}
\label{eq:stoch_SL}
dZ = \left[(\mu + \mathrm{i}\omega_0)Z - (c_3+\mathrm{i}\beta)|Z|^2 Z\right]dt
     + \sqrt{2D}\,dW ,
\end{equation}
where $D$ is the diffusion coefficient representing the intensity of the external noise, and $dW= dW_1 + \mathrm{i}dW_2$ is a complex Wiener process driven by independent standard real-valued Brownian motions, and $c_3$ is a cubic saturation coefficient, retained rather than normalized to unity for the reason listed in the next section. 

Let us write the complex state in Cartesian coordinates, $Z = X_1 + \mathrm{i}X_2$, with $A = |Z| = (X_1^2+X_2^2)^{1/2}$ and $\theta = \arg Z = \operatorname{atan2}(X_2,X_1)$, so that $Z = A e^{\mathrm{i}\theta}$. 

As such, \eqref{eq:stoch_SL} is a two-dimensional It\^o system with additive, isotropic noise,
\begin{equation}
\label{eq:appA_cart}
\begin{aligned}
dX_1 &= F_1(X_1,X_2)\,dt + \sqrt{2D}\,dW_1, &\qquad
F_1 &= \mu X_1 - \omega_0 X_2 - (X_1^2+X_2^2)\,(c_3 X_1 - \beta X_2),\\
dX_2 &= F_2(X_1,X_2)\,dt + \sqrt{2D}\,dW_2, &\qquad
F_2 &= \mu X_2 + \omega_0 X_1 - (X_1^2+X_2^2)\,(c_3 X_2 + \beta X_1),
\end{aligned}
\end{equation}
whose diffusion matrix is $\sigma\sigma^{\!\top} = 2D\,I_2$, constant and state-independent.

For a twice-differentiable test function $\varphi(X_1,X_2)$, It\^o's lemma \cite{oksendal2003stochastic} applied to Eq.~\eqref{eq:appA_cart} gives 
\begin{equation}
\label{eq:appA_ito}
d\varphi = \Big[\nabla \varphi \cdot F \;+\; D\,\nabla^2 \varphi\Big]dt
     \;+\; \sqrt{2D}\;\nabla \varphi \cdot dW ,
\qquad
\nabla^2 \varphi = \partial^2_{X_1} \varphi + \partial^2_{X_2} \varphi ,
\end{equation}
the second-order term reducing to $\tfrac12\operatorname{tr}
(\sigma\sigma^{\!\top}\nabla^2 \varphi) = D\,\nabla^2 \varphi$ by isotropy. The two coordinate functions have
\begin{equation}
\label{eq:appA_grads}
\nabla A = \begin{pmatrix}\cos\theta\\ \sin\theta\end{pmatrix},
\qquad
\nabla^2 A = \frac{1}{A},
\qquad
\nabla \theta = \frac{1}{A}\begin{pmatrix}-\sin\theta\\ \cos\theta\end{pmatrix},
\qquad
\nabla^2 \theta = 0 .
\end{equation}
The two Laplacians are important since $|Z|$ is not harmonic in the plane and picks up a second-order term, whereas $\arg Z$ is harmonic and does not. The first-order terms are the radial and angular components of the drift field. Using $F(Z) = (\mu + \mathrm{i}\omega_0)Z - (c_3+\mathrm{i}\beta)|Z|^2 Z$ and $Z = ae^{\mathrm{i}\theta}$,
\begin{equation}
\label{eq:appA_drift}
\nabla A \cdot F = \operatorname{Re}\!\big[F e^{-i\theta}\big]
= \mu A - c_3 A^3 ,
\qquad
\nabla \theta \cdot F = \frac{1}{A}\operatorname{Im}\!\big[F e^{-i\theta}\big]
= \omega_0 - \beta A^2 .
\end{equation}
The noise terms follow from the gradients in Eq.~\eqref{eq:appA_grads}:
\begin{equation}
\label{eq:appA_noise}
dW_A \;=\; \cos\theta\,dW_1 + \sin\theta\,dW_2 ,
\qquad
dW_\theta \;=\; -\sin\theta\,dW_1 + \cos\theta\,dW_2 ,
\end{equation}
that is, the projections of the planar increment onto the radial and tangential directions. Each is a scalar It\^o integral of a unit vector against a standard planar Brownian motion, so each has quadratic variation
$d\langle W_A\rangle = d\langle W_\theta\rangle = dt$ and is itself a standard Wiener process by L\'evy's characterization; their quadratic covariation vanishes identically, $d\langle W_A, W_\theta\rangle = (-\cos\theta\sin\theta + \sin\theta\cos\theta)\,dt = 0$, so $W_A$ and $W_\theta$ are independent. The rotation is state-dependent, but it is a rotation, and Brownian motion in the plane is invariant under it. Substituting Eqs.~\eqref{eq:appA_grads}-\eqref{eq:appA_noise} into Eq.~\eqref{eq:appA_ito} and noting that $\nabla A$ is a unit vector while $|\nabla\theta| = 1/A$ gives the pair fitted in the main text,
\begin{align}
\label{eq:amp}
dA &= \left(\mu A - c_3 A^3 + \frac{D}{A}\right)dt + \sqrt{2D}\,dW_A ,\\
\label{eq:pha}
d\theta &= \left(\omega_0 - \beta A^2\right)dt + \frac{\sqrt{2D}}{A}\,dW_\theta ,
\end{align}
Setting $D = 0$ recovers the deterministic polar form of Eq.~\eqref{eq:stoch_SL}, confirming that $\mu A - c_3 A^3$ and $\omega_0 - \beta A^2$ are the deterministic drifts and that $D/A$ is the entire stochastic contribution to the amplitude drift. Because the diffusion coefficient of Eq.~\eqref{eq:stoch_SL} is constant and state-independent, its It\^o and Stratonovich readings coincide, so the $D/A$ term of Eq.~\eqref{eq:amp} arises solely from the nonlinear change of variables to polar coordinates and not from the interpretation of the stochastic integral \cite{oksendal2003stochastic}. The term $D/A$ is the characteristic two-dimensional Bessel drift induced by isotropic planar diffusion; the full radial process is therefore a Stuart-Landau radial diffusion with a Bessel-type noise correction \cite{revuz1999continuous}.

The physical interpretations of the model parameters are as follows: $\omega_0$ is the baseline-frequency estimate which is expected to be near 3 cpm for gastric rhythms; $\mu$ is the signed distance to the Hopf bifurcation, dictating rhythm self-sustainment ($\mu > 0$) and structural resilience to perturbations; $c_3$ represents the non-linear cubic saturation; $\beta$ is the nonisochronicity or shear parameter, quantifying how amplitude fluctuations modulate the instantaneous frequency; and $D$ scales the intrinsic stochastic drive of the system.

\runin{Estimating the cubic saturation.} Equation~\eqref{eq:deter_SL} is written in the conventional normal form, with a unit real cubic coefficient. That normalization is reached by rescaling $Z \mapsto \sqrt{c_3}\,Z$, which also divides the shear by $c_3$ and, in the stochastic form of Eq.~\eqref{eq:stoch_SL}, carries the noise intensity to $c_3 D$. Fixing the cubic coefficient at unity in a stochastic model would therefore tie the oscillator's operating radius to the strength of its own noise, and would impose a second scale convention on top of the per-recording median amplitude normalization introduced below to make parameters comparable across recording systems. We therefore retain an explicit cubic saturation coefficient $c_3 > 0$ and estimate it from the data. This preserves their distinct formal roles within the fitted amplitude equation: $\mu$ governs the linear radial drift, $c_3$ the cubic radial term, and $D$ the diffusion scale. Their physiological interpretation is considered separately below.

\runin{Relation between the Stuart-Landau and phase-harmonic models.} The relation between the stochastic Stuart-Landau model and the phase-only analysis used in applied EGG can be made explicit. For $\mu > 0$ and $c_3 > 0$ the deterministic amplitude drift $\mu A - c_3 A^3$ has the nonzero fixed point $a^\ast = \sqrt{\mu/c_3}$, with
\begin{equation}
\left.\frac{d}{dA}\left(\mu A - c_3 A^3\right)\right|_{A = a^\ast}
= \mu - 3c_3 a^{\ast 2} = -2\mu < 0 ,
\end{equation}
so the limit cycle is linearly stable and $\mu$ sets the rate at which amplitude perturbations decay, as stated in the Introduction. Clamping the amplitude at that value, $A(t) \equiv a^\ast$, eliminates Eq.~\eqref{eq:amp} and leaves the phase channel alone:
\begin{equation}
\label{eq:appA_phasediff}
d\theta = \big(\omega_0 - \beta a^{\ast 2}\big)dt
        + \frac{\sqrt{2D}}{a^\ast}\,dW_\theta ,
\end{equation}
a Wiener process with constant drift. Its increments are stationary, and the phase spreads diffusively, $\operatorname{Var}[\theta(t) - \theta(0)] = 2Dt/a^{\ast 2}$, giving a phase diffusion coefficient $D_\theta = D/a^{\ast 2}$ and a coherence time that grows with the squared orbit radius. Equation~\eqref{eq:appA_phasediff} is the generative model implicit in conventional Hilbert-phase EGG analysis: a constant-rate rotation observed through phase diffusion. Setting $D = 0$ in the phase channel as well gives
\begin{equation}
\label{eq:reduction}
\frac{d\theta}{dt} = \omega_0 - \beta a^{\ast 2} = \text{const},
\end{equation}
the isochronous harmonic oscillator with constant amplitude $a^\ast$ and constant frequency. The model class is therefore a strict hierarchy,
\begin{equation}
\underbrace{\text{harmonic}}_{A \equiv a^\ast,\; D = 0}
\;\subset\;
\underbrace{\text{phase-only}}_{A \equiv a^\ast,\; D > 0}
\;\subset\;
\underbrace{\text{Stuart-Landau}}_{A(t)\ \text{free},\; D > 0},
\end{equation}
and each restriction is a testable parameter constraint rather than a change of framework, which is what makes every comparison with conventional practice a nested-model question.

Two properties of the demonstrated amplitude clamp are worth stating explicitly as they emphasize what phase-only analysis can and cannot estimate. First, under the clamp $\omega_0$ and $\beta$ enter Eq.~\eqref{eq:appA_phasediff} only through the composite $\omega_0 - \beta a^{\ast 2}$ and are not separately identifiable: a single dominant-frequency estimate is an estimate of that composite, and no amount of phase data at fixed amplitude can decompose it. Separating the intrinsic rate from the shear requires amplitude to vary. Second, the clamp is exact only if the amplitude is genuinely constant. When it is not, the phase-only model is misspecified by a term that is directly observable: writing $\bar A^2 = \langle A^2 \rangle$,
\begin{equation}
\label{eq:appA_resid}
\dot\theta(t)
= \underbrace{\big(\omega_0 - \beta \bar A^{2}\big)}_{\text{fitted constant}}
\;-\; \beta\big(A^2(t) - \bar A^{2}\big) \;+\; \text{noise} ,
\end{equation}
so the residual of a constant-frequency fit carries the amplitude fluctuation scaled by $-\beta$. 

\runin{Meal state as separate fits.} For the prandial cohort, the fasted and fed blocks of each recording are treated as two separate records: each is preprocessed, band-passed and Hilbert-transformed on its own, each is fit separately, and the resulting parameters are compared within recording as a paired contrast, so that no fit spans the transition. The fasted state is defined as the recording before the meal marker and the fed state is defined as the recording 5 minutes after the meal marker. 

\runin{Baseline-frequency drift and the effective shear.} The estimated gastric-band baseline frequency drifts slowly over multi-hour recordings \cite{ArtifactRejectionMethodology}, and a single global intercept may not track these changes well. In addition, $\beta = -d\dot\theta/d(A^2)$ carries amplitude scale, so an un-normalized $\beta$ shrinks by the $A^2$-expansion factor when a meal makes the carrier amplitude larger, and falsely appears more isochronous. We therefore replace the global intercept with a slow local baseline and regress local deviations of a median-normalized amplitude, with no free intercept:
\begin{equation}
\label{eq:beta_eff}
\left[\dot\theta(t) - \omega_0(t)\right]
 = -\beta_{\mathrm{eff}}\left[\tilde A^2(t) - \langle\tilde A^2\rangle_t\right] + \varepsilon(t),
\qquad \tilde A = A/\mathrm{median}(A),
\end{equation}
where $\omega_0(t)$ and $\langle\tilde A^2\rangle_t$ are 5-minute moving averages. For the overnight cohort, whose length permits cross-validation, these are strictly \emph{causal} trailing averages taken over samples at or before $t$ only, so that no held-out row's baseline can depend on its own future. The prandial fits report no held-out metric and use a centered window, which is better conditioned on 30-60\,minute records; the consequences of that choice are quantified in Section~\ref{sec:results}. 

We define $\beta_{\mathrm{eff}}$ as a relative, dimensionless effective shear, and a block-bootstrap standardization $z_\beta = \beta_{\mathrm{eff}}/\mathrm{SE}$ places estimates on a common standardized scale across cohorts. Both terms are high-pass operations on quantities that are themselves slowly varying, so the regression uses only the fast, within-window covariation of phase velocity with squared amplitude and discards the slow common drift that a global intercept would otherwise absorb. 
 
\subsection{Model parameter estimation}
\runin{Model estimation.} The Euler-Maruyama discretization is a numerical method used to approximate solutions to stochastic differential equations, where it extends the standard forward Euler method from ordinary differential equations by adding a random noise term scaled to a discrete time step $\Delta t$ and Brownian motion increment $\Delta W$ \cite{NumericalSolutionStochastic}. For a general It\^o stochastic differential equation of the form:

$$dZ_{t}=f(Z_{t},t)\,dt+g(Z_{t},t)\,dW_{t}$$

The Euler-Maruyama update scheme divides time into steps $t_n = n\Delta t$ and updates the approximation $Z_n \approx Z(t_n)$ via the iterative formula

$$Z_{n+1}=Z_{n}+f(Z_{n},t_{n})\,\Delta t+g(Z_{n},t_{n})\,\Delta W_{n}$$
where $\Delta t = t_{n+1} - t_n$ is the deterministic step size and $\Delta W_n = W_{t_{n+1}} - W_{t_n} = Q_n \sqrt{\Delta t}$ represents the Brownian motion increment, with $Q_n \sim \mathcal{N}(0, 1)$ being a standard normal random variable. Under this discretization with additive noise, the one-step transition density is Gaussian and both drifts are linear in the unknown parameters, so maximization of the likelihood separates and the estimator is available in closed form as weighted least squares:

\begin{itemize}
\item {\bf Amplitude}: The Euler-Maruyama update scheme for \eqref{eq:amp}
results in regression amplitude response variables $Y_i=(A_{i+1}-A_i)/\Delta t$
\begin{equation}
Y_i = \begin{bmatrix} A_i & -A_i^3 & 1/A_i\end{bmatrix}
      \begin{bmatrix}\mu \\ c_3 \\ D\end{bmatrix} + \varepsilon_i ,
\end{equation}

\item {\bf Phase}: 
The Euler-Maruyama update scheme for \eqref{eq:pha}
results in regression phase response variables $Y_i$ given by $Y_i = \mathrm{wrap}(\theta_{i+1}-\theta_i)/\Delta t$
\begin{equation}
Y_i = \begin{bmatrix} 1 & -A_i^2 \end{bmatrix}
      \begin{bmatrix}\omega_0 \\ \beta\end{bmatrix} + \varepsilon_i .
\end{equation}
\end{itemize}
Stacking every sample of a channel gives the response vector $\mathbf{Y}_\bullet$, the design matrix $\mathbf{X}_\bullet$ whose $i$-th row is $\mathbf{x}_i^\top$, and the parameter vector $\boldsymbol\eta_\bullet$, where $\bullet \in \{\mathrm{amp},\mathrm{pha}\}$ indexes the channel: $\boldsymbol\eta_{\mathrm{amp}} = (\mu, c_3, D)^\top$ with $\mathbf{X}_{\mathrm{amp}} \in \mathbb{R}^{n \times 3}$, and $\boldsymbol\eta_{\mathrm{pha}} = (\omega_0, \beta)^\top$ with $\mathbf{X}_{\mathrm{pha}} \in \mathbb{R}^{n \times 2}$. The weighted least-squares estimate of either channel is
\begin{equation}
\label{eq:wls}
\hat{\boldsymbol\eta}_\bullet
  = \bigl(\mathbf{X}_\bullet^\top \mathbf{K}_\bullet \mathbf{X}_\bullet\bigr)^{-1}
    \mathbf{X}_\bullet^\top \mathbf{K}_\bullet \mathbf{Y}_\bullet ,
\qquad
\mathbf{K}_\bullet = \operatorname{diag}(k_1,\dots,k_n) ,
\qquad
k_i = k_i^{\mathrm{den}}\,k_i^{\mathrm{var}} ,
\end{equation}
from which $\mu$, $c_3$, $D$, $\omega_0$, $\beta$ and $a^\ast = \sqrt{\mu/c_3}$ follow directly.

In the polar parameterization all drift parameters are linear, so the fit reduces to weighted least squares. The amplitude channel is a single closed-form solve, and the phase channel is a short iteratively reweighted sequence of the same linear solve. The increment-variance model follows from the Euler-Maruyama discretization. The diagonal weight $k_i$ of Eq.~\eqref{eq:wls} combines a state-density weight $k_i^{\mathrm{den}}$, which prevents sparsely visited amplitudes from dominating, with a pointwise-variance (generalized least squares) weight $k_i^{\mathrm{var}}$, since the increment variance of Eq.~\eqref{eq:pha} scales as $1/A^2$ and low-amplitude samples are correspondingly less informative about the phase drift.

To account for phase slips, which may occur when there are low signal-to-noise amplitude drops \cite{gogoiPhaseSlipsCoupled2023,ramonPhaseSlipsExtracted2025}, the phase channel is fit by the Huber M-estimator \cite{huberRobustEstimationLocation1964}, solved by iteratively reweighted least squares (IRLS). Beginning from the weighted least-squares solution, at each iteration the residuals $R_i = Y_i - \mathbf{x}_i^{\top}\hat{\boldsymbol\eta}_{\mathrm{pha}}$ are scaled by a robust estimate of their spread, $\hat\sigma_R = 1.4826\,\mathrm{MAD}(R)$, where $\mathrm{MAD}$ means mean absolute deviation, and each sample receives a Huber weight
\begin{equation}
\label{eq:huber}
h_i =
\begin{cases}
1, & |R_i/\hat\sigma_R| \le c,\\[2pt]
c \,/\, |R_i/\hat\sigma_R|, & |R_i/\hat\sigma_R| > c,
\end{cases}
\qquad c = 1.345 ,
\end{equation}
so residuals inside the threshold keep unit weight while tail residuals, in this case the low-amplitude phase-slip outliers, are down-weighted in proportion to how far they lie in the tail. The constant $c = 1.345$ yields 95\% efficiency relative to least squares under Gaussian residuals.

The robustness weights enter the estimator of Eq.~\eqref{eq:wls} by multiplying the diagonal, so that each IRLS iteration is the same weighted linear solve with an updated weight matrix. Writing $m$ for the iteration index, and initializing $\hat{\boldsymbol\eta}_{\mathrm{pha}}^{(0)}$ at the weighted least-squares solution of Eq.~\eqref{eq:wls},
\begin{equation}
\label{eq:irls}
R_i^{(m)} = Y_i - \mathbf{x}_i^{\top}\hat{\boldsymbol\eta}_{\mathrm{pha}}^{(m)} ,
\qquad
\mathbf{K}^{(m)} = \operatorname{diag}\!\bigl(k_i^{\mathrm{den}}\,k_i^{\mathrm{var}}\,h_i^{(m)}\bigr) ,
\qquad
\hat{\boldsymbol\eta}_{\mathrm{pha}}^{(m+1)}
  = \bigl(\mathbf{X}_{\mathrm{pha}}^\top \mathbf{K}^{(m)} \mathbf{X}_{\mathrm{pha}}\bigr)^{-1}
    \mathbf{X}_{\mathrm{pha}}^\top \mathbf{K}^{(m)} \mathbf{Y}_{\mathrm{pha}} ,
\end{equation}
where $h_i^{(m)}$ is Eq.~\eqref{eq:huber} evaluated at $R_i^{(m)}/\hat\sigma_R$. The robust scale $\hat\sigma_R$ is recomputed from the current residuals at each iteration, and the recursion is run until the relative change in $\hat{\boldsymbol\eta}_{\mathrm{pha}}$ falls below $10^{-6}$ or 50 iterations are reached. Because only the diagonal changes between iterations, the down-weighting acts on the estimate exactly as a change of observation weights: a sample whose residual falls in the tail contributes to the normal equations in proportion to $h_i < 1$ rather than being removed.

At a fixed point of Eq.~\eqref{eq:irls} the estimate satisfies the Huber estimating equation,
\begin{equation}
\label{eq:mest}
\sum_{i=1}^{n} k_i^{\mathrm{den}}\,k_i^{\mathrm{var}}\;
\psi_c\!\left(\frac{R_i}{\hat\sigma_R}\right)\mathbf{x}_i \;=\; \mathbf{0} ,
\qquad
\psi_c(u) = \max\{-c,\ \min(c,u)\} ,
\end{equation}
so the reweighted solve is the M-estimator of \cite{huberRobustEstimationLocation1964} under the same density and variance weights, rather than an ad hoc attenuation of large residuals. Equation~\eqref{eq:mest} reduces to the weighted normal equations of Eq.~\eqref{eq:wls} when every $|R_i| \le c\,\hat\sigma_R$, so the robust and non-robust fits coincide on records without slips. Only the phase channel is reweighted, so the reweighting changes $\hat\omega_0$ and $\hat\beta$, and through Eq.~\eqref{eq:beta_eff} the reported $\beta_{\mathrm{eff}}$, while $\hat\mu$, $\hat c_3$ and $\hat D$ are those of the single amplitude-channel solve and are unaffected. Every sample is retained: no interpolation or deletion of the phase series is performed-so the estimator is robust to the slips without discarding the data that carries the drift. The amplitude channel, whose increments are not subject to these slips, uses the density- and variance-weighted least squares of Eq.~\eqref{eq:wls} without the robustness reweighting.

\runin{Recovery of the effective shear in simulated data.} The quantity validated by simulation is the coefficient reported in the empirical analyses, rather than the physical Stuart-Landau shear $\beta$ of Eq.~\eqref{eq:pha}. Because band-limiting is unavoidable in EGG and alters the physical scale of raw shear, we defined the recovery target as the same local, median-normalized coupling observable estimated by Eq.~\eqref{eq:beta_eff}, evaluated on the known latent state of the simulation:
\begin{equation}
\label{eq:beta_star}
\beta^{\ast}_{\mathrm{eff}} \;=\; \text{slope of }
\left[\dot\theta(t)-\omega_0(t)\right] \text{ on }
-\left[\tilde A^2(t)-\langle\tilde A^2\rangle_t\right],
\qquad
\tilde A(t)=\frac{A(t)}{\mathrm{median}[A(t)]},
\end{equation}
where $\langle\tilde A^2\rangle_t$ denotes the centered 5-minute local mean used in Eq.~\eqref{eq:beta_eff}. The target $\beta^{\ast}_{\mathrm{eff}}$ was computed from the simulator's known latent amplitude $A(t)$ and phase $\theta(t)$ using the same median normalization, local baseline, and Huber regression used for empirical data. Thus, the mapping
$\beta\rightarrow\beta^{\ast}_{\mathrm{eff}}$ tests whether the effective-shear definition represents the imposed phase-amplitude coupling, whereas $\beta^{\ast}_{\mathrm{eff}}\rightarrow\hat{\beta}_{\mathrm{eff}}$ tests recovery through the complete EGG observation and analysis chain.

We generated a hybrid ground-truth process in which amplitude was an Ornstein-Uhlenbeck process parameterized to match the temporal scale and coefficient of variation of the observed EGG envelope, while phase followed the Stuart-Landau phase equation unchanged, including the $-\beta A^2$ drift and $\sqrt{2D}/A$ stochastic term. The real-valued observation was defined as $S(t)=\mathrm{Re}\{Z(t)\}$. To emulate the measurement setting, we added baseline drift and respiratory and cardiac components outside the gastric band, consistent with the decomposition in Eq.~\eqref{eq:decomp}. The resulting signal was passed through every model fitting step and refitted.

Recovery was evaluated by four pre-specified quantities: (i) calibration of $\hat{\beta}_{\mathrm{eff}}$ against $\beta^{\ast}_{\mathrm{eff}}$, summarized by intercept, slope, and $R^2$ with bootstrap confidence intervals; (ii) the proportion of nonzero simulated couplings for which the fitted estimate recovered the correct sign; (iii) the distribution of $\hat{\beta}_{\mathrm{eff}}$ under exactly zero imposed coupling; and (iv) bias and root-mean-square error as functions of record duration. Spearman rank correlation across the simulated coupling grid was also reported because the meal-state and diagnostic claims are ordinal.

Three controls tested estimator specificity. First, under simulated baseline-frequency drift, the local-baseline estimator was compared with a conventional global-intercept regression. Second, the observed signal was multiplied by several constant gains to test whether median normalization made $\hat{\beta}_{\mathrm{eff}}$ invariant to observation scale. Third, the circular-shift, amplitude-adjusted Fourier, and matched autoregressive amplitude surrogates used in the empirical analysis were applied to simulated records with known nonzero coupling; each was expected to eliminate the fitted coupling by destroying the amplitude--phase correspondence while retaining specified properties of the amplitude series.

\subsection{Model application to cutaneous EGG recordings in human subjects}
\runin{Standard errors and conditional-drift agreement metric.} Standard errors are reported by moving-block bootstrap over 400 resamples, with the block length set to 20 seconds, approximately one gastric cycle, so that within-cycle dependence is preserved. Heteroskedasticity- and autocorrelation-consistent standard errors \cite{SimplePositiveSemiDefinite} are computed alongside it as an agreement check.

We assess drift agreement through binned estimates of the conditional mean increment \cite{siegertAnalysisDataSets1998}. Let $U_i$ denote the state of sample $i$ (the amplitude $\tilde A_i$ for Eq.~\eqref{eq:amp}, the squared amplitude $\tilde a_i^2$ for Eq.~\eqref{eq:pha}) and $Y_i$ the corresponding observed increment. The state range is divided into $B = 18$ equal-width bins spanning the 0.5th to 99.5th percentile of $u$, giving edges $e_0 < \dots < e_B$ and bins $\mathcal{B}_b = \{i : e_{b-1} \le U_i < e_b\}$ with counts $n_b = |\mathcal{B}_b|$. Let $\bar{Y}_b$ denote the mean observed increment in bin $b$ and $\overline{\widehat f}_b$ the mean fitted conditional drift over rows in that bin:
\begin{equation}
\label{eq:binmeans}
\bar Y_b = \frac{1}{n_b}\sum_{i \in \mathcal{B}_b} Y_i ,
\qquad
\overline{\widehat f}_b= \frac{1}{n_b}\sum_{i \in \mathcal{B}_b} \hat f_i ,
\end{equation}
the agreement metric is the count-weighted coefficient of determination of the binned conditional mean,
\begin{equation}
\label{eq:r2bin}
R^{2}_{\mathrm{bin}}
= 1 - \frac{\displaystyle\sum_{b \in \mathcal{B}} w_b\,\bigl(\bar Y_b - \hat Y_b\bigr)^{2}}
           {\displaystyle\sum_{b \in \mathcal{B}} w_b\,\bigl(\bar Y_b - \bar Y_{\mathcal{B}}\bigr)^{2}} ,
\qquad
w_b = \frac{n_b}{\sum_{b' \in \mathcal{B}} n_{b'}} ,
\qquad
\bar Y_{\mathcal{B}} = \sum_{b \in \mathcal{B}} w_b\,\bar Y_b ,
\end{equation}
where $\mathcal{B} = \{b : n_b \ge 5\}$ retains only bins with at least five samples and the metric is left undefined unless at least three bins are retained. Bin edges are held fixed at the values obtained from the full record, so held-out halves and bootstrap resamples are scored on the same grid. We report $R^{2}_{\mathrm{bin}}$ rather than the per-sample $R^2$ as the drift is a small deterministic push on large per-sample stochastic variability, so it is strong in the conditional mean and negligible per sample. Model generalization is tested by split-half cross-prediction: the model is fit on the first half of a recording's rows and scored on the second, then fit on the second and scored on the first, and the two held-out halves are pooled. A moving-average local baseline makes the choice of validation scheme consequential, because a held-out row's baseline is itself a filter of neighboring rows. We therefore adopt the scheme under which that dependence is weakest: a single split rather than $k$ interleaved blocks, so that a held-out row is adjacent to training data on one side only, together with the strictly causal trailing average of Eq.~\eqref{eq:beta_eff}, so that no held-out row's baseline depends on its own future.

\runin{Surrogate nulls.} For the model application to the sleep cohort, a family of nulls test whether the amplitude-to-frequency coupling is real or an artifact of the estimator. For each recording, the amplitude time series is replaced by three null distributions that break the temporal correspondence between $A(t)$ and $\dot\theta(t)$ while preserving the amplitude series' own autocorrelation: i) a large circular time shift, which preserves the autocorrelation and marginal distribution exactly; ii) an amplitude-adjusted Fourier transform, which matches the spectrum and the marginal; and iii) an independent stable autoregressive process of order 60 fitted to the real amplitude by the Yule-Walker equations \cite{kreissDetectingPeriodicityGeneral}, which captures the parametric autocorrelation structure and marginal variance of the empirical series. If the fit of Eq.~\eqref{eq:beta_eff} depended on any of the preprocessing steps performed for the model fit rather than on genuine coupling, these coupling-broken series would reproduce it. 

\runin{Use of standardized shear for prandial cohort.} Unlike the overnight sleep cohort, which is a single acquisition system, the prandial cohort pools three: a 25-channel, an 8-channel and a 16-channel ambulatory system whose amplitude calibrations differ by orders of magnitude. Because $\beta_{\mathrm{eff}}$ is defined on the median-normalized amplitude of the recording it is fitted to, it retains that device-dependent scale. Dividing each recording by its own block-bootstrap standard error removes the amplitude scale and weights each recording by how well its own shear is determined. We therefore report $z_\beta$ for the prandial cohort. 

\runin{Comparison of phase-shear measure against high-resolution multi-electrode array features.} To determine whether a single-electrode phase-shear measure carries comparable diagnostic information to array-derived features, we computed multi-electrode phase information as described in \cite{gharibansSpatialPatternsHighResolution2019c}. Features were computed via the phase-gradient method, where each electrode is band-pass filtered to the gastric band, the Hilbert analytic phase is taken at every site, and the spatial phase gradient is formed by finite differences along each grid axis after unwrapping along that axis. From this gradient we derived the propagation direction and the phase-gradient directionality (PGD), which measures how well the gradients align across the array, and, together with the temporal phase derivative, the propagation speed. A sustained wave required a PGD above 0.5 and a speed below 10\,mm\,s$^{-1}$ for at least two consecutive seconds. We performed this comparison on the 7 healthy controls from the 25-channel array, the 5 controls from the 16-channel array, and 24 patients with gastric dysfunction; the remaining 4 controls used a data acquisition system that did not allow for a PGD calculation. All features are computed on a common central $4\times4$ electrode aperture at 2\,cm spacing. Seven array-derived measures were scored as single features against pooled gastric dysfunction, alongside $|z_\beta|$ on the same participants (Table~\ref{tab:array}). These array-derived measures were speed variability, wave propagation speed, \% abnormal wave direction, inter-electrode phase-locking value, directional dispersion, phase-gradient directionality (PGD), and \% sustained waves. Speed variability is the coefficient of variation of the instantaneous propagation speed: s, the standard deviation of the speed estimates across all sustained-wave time points in a block divided by their mean. A wave that travels at a steady speed throughout the block therefore scores low, while one whose speed fluctuates from cycle to cycle scores high, independently of how fast the wave travels on average.

\runin{Statistical methods.} Our unit of inference is based on individual participants. As the model fit is performed on a single recording, and per-channel and per-window values are pooled within that recording rather than counted as separate observations. Because both cohorts contain repeated sessions from the same individuals, every reported statistic first collapses a participant's recordings to a single value, defined as the within-participant median of the recording-level quantity. The sleep cohort comprises 144 recordings from 60 participants, with up to three nights per participant, and the healthy prandial set comprises 26 recordings from 16 participants, with one individual contributing six sessions. The patient arm of the prandial protocol contains one recording per participant, yielding 24 independent recordings. Cohort summaries are therefore medians and interquartile ranges across participants (60 sleep, 16 healthy prandial, and 24 patients). Reported proportions are the within-participant fraction averaged across participants. Confidence intervals on cohort medians are obtained by bootstrap resampling clustered on participant, in which participants are sampled with replacement together with all of their recordings, so that the interval reflects the number of individuals rather than the larger number of correlated recordings.

Paired contrasts between meal states use the Wilcoxon signed-rank test and a paired $t$-test applied to the within-participant median of each participant's fasted-to-fed differences, which defines the paired estimand at this unit of inference. Between-group comparisons use pairwise Mann-Whitney tests with Holm correction and Cliff's $\delta$ as the effect size, which is appropriate for the small and unequal group sizes.

Discriminative performance in Fig.~\ref{fig:fig4}D is summarized by the area under the receiver operating characteristic curve. Ninety-five-percent confidence intervals are obtained from 10,000 stratified bootstrap resamples, and two-sided significance is assessed using the resampled permutations. Leave-one-out cross-validation is also performed, with the Youden threshold re-estimated on the remaining participants at each iteration, to assess whether performance is driven by any single participant. Because the seven array-derived measures constitute one family of tests evaluated on the same cohort, their permutation $p$ values are corrected across the seven measures using the Holm-Bonferroni method; both uncorrected and corrected values are reported. The $|z_\beta|$ measure was specified in advance as the primary marker and is therefore reported without this correction.

Each array-derived feature is additionally compared with $|z_\beta|$ using a bootstrap distribution of the difference in AUC (Fig.~\ref{fig:fig4}E). Within each of 10,000 resamples, participants are sampled with replacement separately from the dysfunction and control groups, and both markers are evaluated on the same resampled participants. The difference between their AUCs is then recorded. Pairing the resample in this way removes sampling variation shared by the two markers, which would not be accounted for by comparing their marginal confidence intervals. The reported interval is the 2.5th to 97.5th percentile of the resulting distribution, and an interval containing zero indicates that the two markers are not distinguishable at this sample size.

\section{Experiments}
\label{sec:results}

\subsection{Parameter recovery in simulated data}
\label{sec:sim}
We used ground-truth simulations to establish which of the two reported parameters our band-limited measurement can estimate. The intrinsic frequency $\omega_0$ is reported in physical units and therefore requires conventional recovery. The effective shear $\beta_{\mathrm{eff}}$ is reported only through contrasts, and requires a recovery test built around the estimand the empirical analysis actually uses rather than around the physical Stuart-Landau coefficient.

\begin{figure}[!ht]
\centering
 \includegraphics[width=\textwidth]{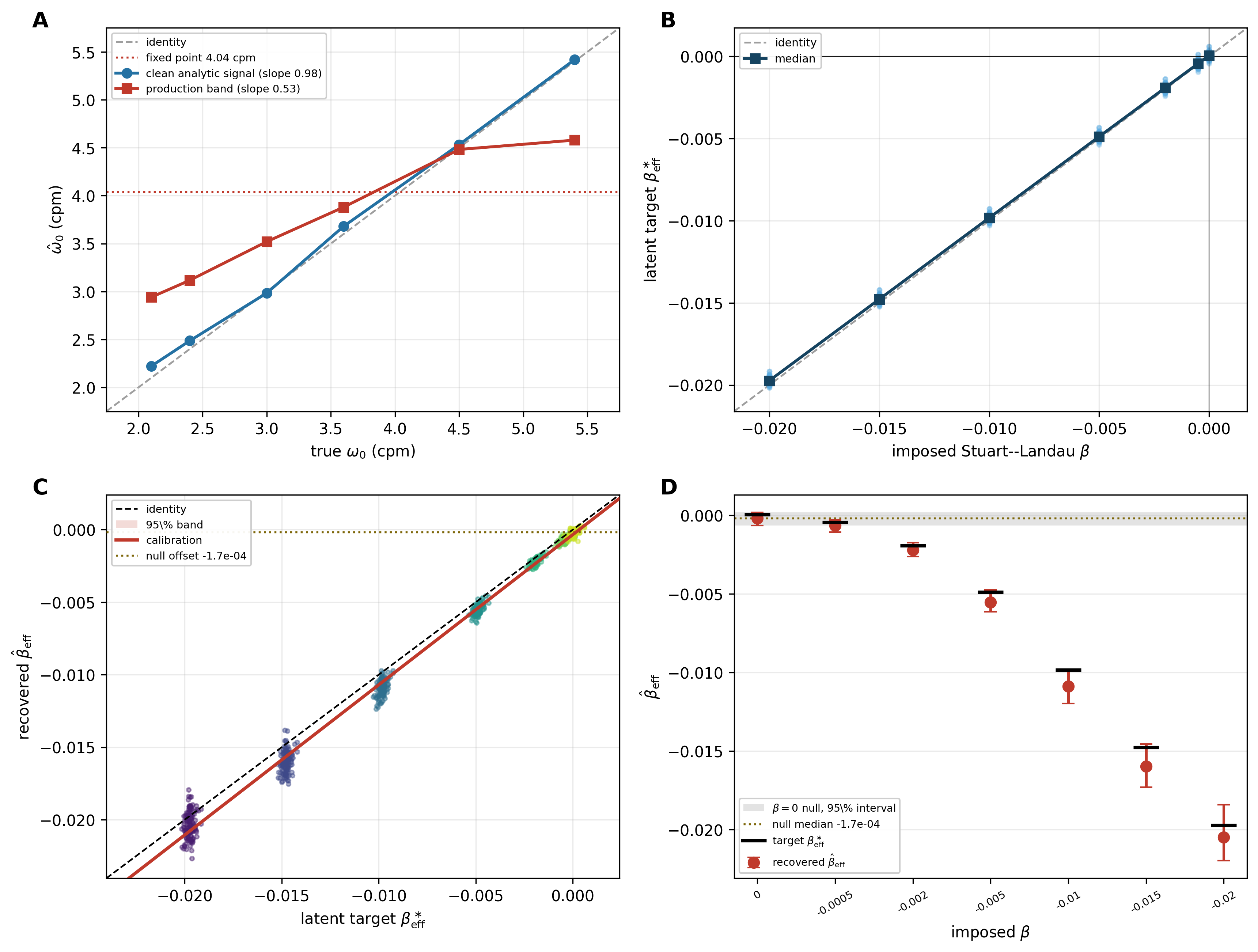}
\caption{\textbf{Parameter recovery in simulated data.}  ({A}) Recovery of the intrinsic frequency across ground-truth carrier frequencies spanning the gastric range. From the clean analytic signal $\hat\omega_0$ is near-identity (slope 0.98); after the band-pass the estimate is shrunk toward a fixed point. ({B}) Validity of the estimand. The latent effective-shear target $\beta^{\ast}_{\mathrm{eff}}$, obtained by evaluating the local-baseline, median-normalized regression functional on the known latent amplitude and phase, is essentially proportional to the imposed Stuart-Landau shear $\beta$ (slope 0.989, $r = 1.000$), so the effective-shear definition itself introduces negligible bias. Light points, individual replicates; squares, medians. ({C}) Recovery through the complete pipeline over the negative range the empirical recordings occupy. All 840 replicates are shown, colored by imposed $\beta$, with the fitted calibration (solid) and its 95\% band, the identity line (dashed), and the estimated null offset (dotted). Calibration slope 1.036 (95\% CI [1.028, 1.043]), intercept $-3.4\times10^{-4}$, $R^2 = 0.994$, Spearman $\rho = 0.987$, and the sign of the coupling recovered in 99.7\% of the 720 nonzero replicates. ({D}) Dose response and null floor. Points and intervals give the median and 2.5-97.5 percentile of $\hat\beta_{\mathrm{eff}}$ at each imposed level; black bars mark the corresponding latent targets $\beta^{\ast}_{\mathrm{eff}}$. Recovery is monotone, and the shaded band and dotted line give the 95\% interval and median of the estimates when the imposed coupling is exactly zero, showing a small negative pipeline offset ($-1.7\times10^{-4}$).}
\label{fig:fig1}
\end{figure}

\runin{Intrinsic frequency is recoverable.} Sweeping the simulated carrier across the gastric range, from 2.1 to 5.4\,cpm, the clean analytic signal returns $\hat\omega_0$ close to the identity relation (slope 0.98, $r = 0.98$; Fig.~\ref{fig:fig1}A). Production filtering does not add a constant offset but shrinks the estimate toward a fixed point, with fitted relation $\hat\omega_0 = 0.53\,\omega_0 + 1.90$, a pull of $k = 0.47$ toward 4.04\,cpm, the frequency at which $\hat\omega_0 = \omega_0$. The resulting bias is therefore signed and depends on the carrier: $+0.84$\,cpm at a 2.1\,cpm truth, $+0.52$ at 3.0\,cpm, $-0.02$ at 4.5\,cpm and $-0.82$ at 5.4\,cpm. At the empirical median of 3.16\,cpm the relation implies an upward bias of approximately 0.4\,cpm under the simulated operating condition. We therefore report $\omega_0$ as a filter-conditioned estimate of the baseline-frequency rather than an exactly calibrated one. The empirical distribution is nevertheless not determined by the filter: its median lies 0.88\,cpm below the fixed point and retains between-recording variation, whereas a filter-dominated estimate would return 4.04\,cpm in every recording.

\runin{The simulated envelope matches the recordings.} The observed post-pipeline envelope has an autocorrelation time of 26\,seconds and a coefficient of variation of 0.66. A stationary white-noise-driven Stuart-Landau amplitude equation reproduces neither at any operating point examined (7.5\,s, 0.43), its envelope being dominated by rapid stochastic fluctuation rather than the slow structured modulation present in the EGG, which is why the hybrid generator of Section~\ref{sec:methods} was used; its envelope matches the recordings on both quantities.

\runin{The estimand is unbiased and the pipeline recovers it.} At the latent level the effective-shear target was essentially proportional to the imposed coefficient, with $\beta^{\ast}_{\mathrm{eff}}$ regressed on $\beta$ giving slope 0.989 and
$r = 1.000$ (Fig.~\ref{fig:fig1}B), so the local-baseline definition introduces negligible bias under envelope dynamics matched to the recordings. Recovery was then assessed over the negative range the empirical recordings occupy. In that range $\hat\beta_{\mathrm{eff}}$ recovers $\beta^{\ast}_{\mathrm{eff}}$ with calibration slope 1.036 (95\% CI [1.028, 1.043]), intercept $-3.4\times10^{-4}$ and $R^2 = 0.994$, with Spearman $\rho = 0.987$ and correct sign recovery in 99.7\% of the 720 nonzero replicates (Fig.~\ref{fig:fig1}C). Recovery is monotone with no reversals across the grid, including at the weak-effect end: imposed couplings of $-0.020$, $-0.015$, $-0.010$, $-0.005$, $-0.002$ and $-0.0005$ were recovered as $-0.0205$, $-0.0160$, $-0.0109$, $-0.0055$, $-0.0022$ and $-0.0007$ respectively (Fig.~\ref{fig:fig1}D).

\runin{A null floor bounds the interpretation of an individual estimate.} When the imposed
coupling is exactly zero the estimator returns a small negative value, median $-1.7\times10^{-4}$, with 80\% of replicates negative. The sign of a single estimate is therefore not by itself evidence of amplitude-frequency coupling, and we interpret the raw $\beta_{\mathrm{eff}}$ as a relative index rather than a physical coefficient to be tested against zero near that floor; the surrogate-standardized $z_\beta$ is retained for zero-referenced inference. Taken together, these simulations establish that under envelope dynamics matched to the EGG recordings, the pipeline recovers the direction, ordering and relative magnitude of negative amplitude-frequency coupling, which is the regime the empirical EGG data falls into.

\begin{figure}[!ht]
\centering
\includegraphics[width=\textwidth]{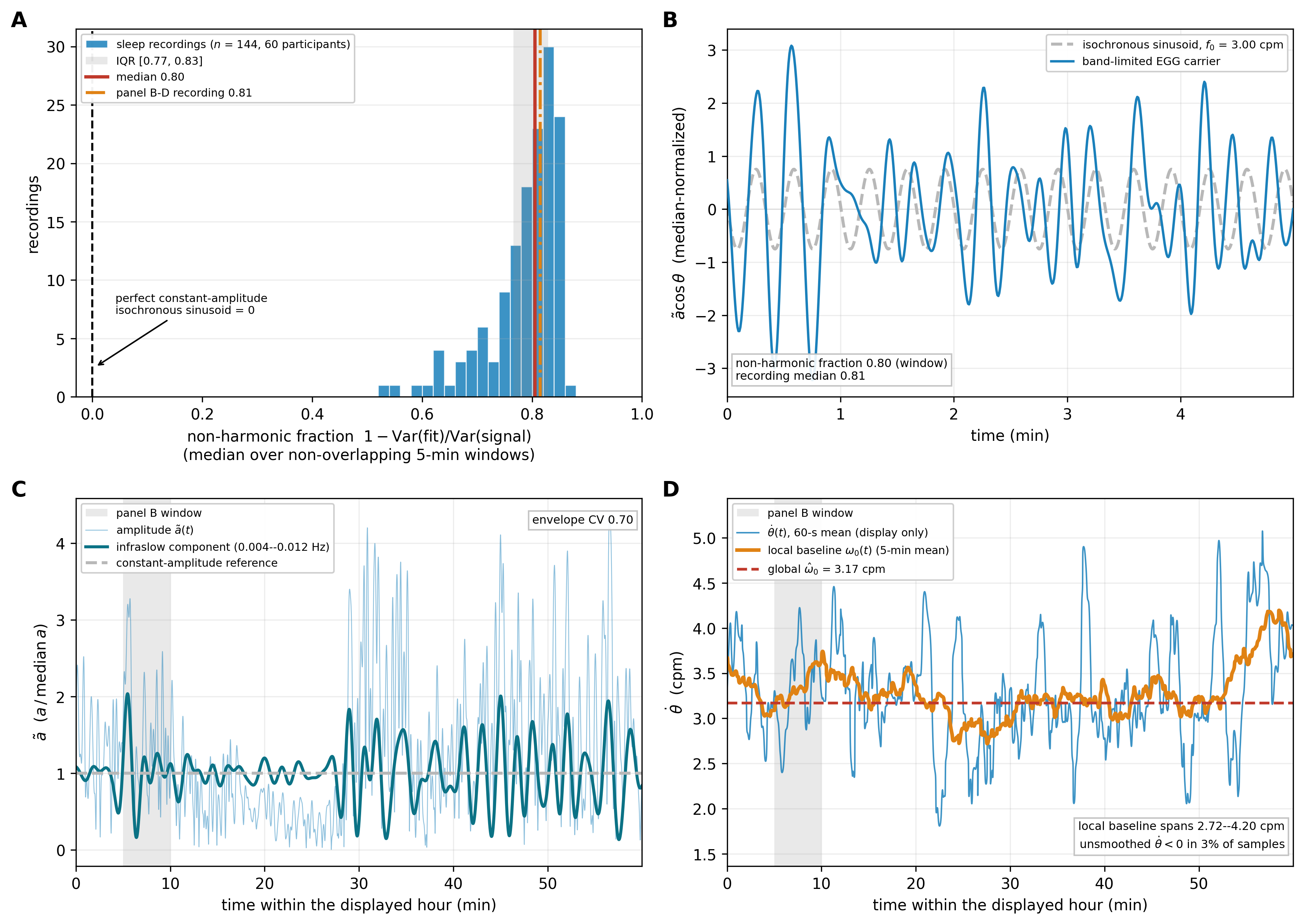}
\caption{\textbf{The overnight EGG carrier is not an isochronous harmonic oscillator.} ({A}) Cohort distribution of the non-harmonic fraction over the sleep cohort (144 recordings from 60 participants); the median and IQR marked are the repeated-measures values across participants, 0.80 with IQR [0.77, 0.83], while the histogram shows every recording. Approximately $\frac{4}{5}$th of the gastric-band signal's variance lies outside the reduced model of Eq.~\eqref{eq:reduction} in a typical EGG recording. The recording used in B-D is marked and sits at the cohort median. ({B}) The gastric-band EGG signal $\tilde A\cos\theta$ (blue) is plotted over a 5-minute window against the least-squares isochronous sinusoid at that window's own spectral peak (grey). ({C}) The amplitude $\tilde A(t)$ over the surrounding hour against the constant-amplitude reference $\tilde a \equiv 1$ the reduced model assumes (grey), with the infraslow (0.004-0.012\,Hz) fluctuation of the amplitude envelope superposed on that reference. ({D}) The instantaneous frequency over the same hour with the centered 5-minute local baseline $\omega_0(t)$ of Eq.~\eqref{eq:beta_eff} and the single global $\hat\omega_0 = 3.17$\,cpm a conventional fit assigns to the whole recording.}
\label{fig:fig2}
\end{figure}
\subsection{Application to overnight sleep cohort}
\runin{The EGG signal is not an isochronous harmonic oscillator.} We first ask whether the constant-amplitude, constant-frequency description implicit in conventional analysis is adequate. Each band-limited segment was compared with the least-squares isochronous sinusoid at that segment's own spectral peak, that is, with the best-fitting member of a harmonic oscillator model of Eq.~\eqref{eq:reduction}, and the departure was quantified as the non-harmonic fraction $1 - \mathrm{Var}(\mathrm{fit})/\mathrm{Var}(\mathrm{signal})$, which would be zero for a perfect constant-amplitude sinusoid at a single fixed frequency. This measure was computed in non-overlapping 5-minute windows and summarized as the median across windows within a recording. The harmonic oscillator model accounts for only a minority of the EGG signal's variance. The non-harmonic fraction has cohort median 0.80 (IQR [0.77, 0.83]) across the 60 sleep participants (Fig.~\ref{fig:fig2}A), and a representative window against its own best-fitting isochronous sinusoid is shown in Fig.~\ref{fig:fig2}B. We demonstrate this behavior in one representative participant where $A(t)$ fluctuates over time (Fig.~\ref{fig:fig2}C), and $\dot\theta(t)$ departs from a flat $f_0$ by enough that its own slow baseline spans 2.7 to 4.2\,cpm within a single hour (Fig.~\ref{fig:fig2}D). The amplitude channel is thus carrying variance that the phase-only reduction discards by construction, which is the motivation for fitting both amplitude and phase channels in Eq. \ref{eq:amp}.
\begin{table}[ht]
\centering\small
\begin{tabular}{@{}llrrr@{}}
\toprule
Parameter & & Median [IQR] & Median SE & Recordings with \\
          & &              &           & $|\hat\theta| > 2\,\mathrm{SE}$ \\
\midrule
\multicolumn{5}{@{}l}{\textit{Phase channel}}\\
$\omega_0$        & intrinsic rate (cpm)          & 3.144 [3.054, 3.235]          & --      & --      \\
$\omega_0$        & intrinsic rate (rad\,s$^{-1}$) & 0.3292 [0.3198, 0.3387]      & 0.0032  & 100\%   \\
$\beta$           & shear                          & $-0.0049$ [$-0.0069$, $-0.0031$] & 0.0008 & 96\%   \\
$\beta_{\mathrm{eff}}$ & local effective shear     & $-0.0064$ [$-0.0091$, $-0.0044$] & --   & 100\%$^{\dagger}$ \\
\addlinespace
\multicolumn{5}{@{}l}{\textit{Amplitude channel}}\\
$\mu$             & growth rate (s$^{-1}$)         & $+0.0001$ [$-0.0009$, $+0.0012$] & 0.0013 & 2\%   \\
$c_3$             & cubic saturation               & 0.00042 [0.00032, 0.00057]    & 0.00026 & 41\%  \\
$D$               & diffusion                      & 0.00061 [0.00008, 0.00115]    & 0.00009 & 85\%  \\
\addlinespace
\multicolumn{5}{@{}l}{\textit{Conditional-drift agreement metric}}\\
$R^2_{\mathrm{bin}}$ & amplitude                   & 0.862 [0.657, 0.935]          & --      & --      \\
$R^2_{\mathrm{bin}}$ & phase, local $\omega_0(t)$   & 0.675 [0.556, 0.756]          & --      & --      \\
\bottomrule
\end{tabular}
\caption{\textbf{Stuart-Landau parameter estimates for the overnight sleep cohort.} Stuart-Landau parameter estimates for the overnight sleep cohort ($n = 60$ participants, 144 recordings, up to three nights each), reported as a repeated measure: each participant contributes the median of their own nights, and the table gives the median [IQR] across the 60 participants (Section~\ref{sec:methods}). Amplitude is median-normalized, so $c_3$, $D$ and $\beta$ are dimensionless. The median SE is the median moving-block bootstrap standard error (400 resamples), aggregated the same way. The last column is the within-participant fraction of recordings whose estimate exceeds twice its own standard error, averaged over participants; $^{\dagger}$ for $\beta_{\mathrm{eff}}$, which carries no comparable standard error, the fraction of recordings with a negative estimate is given instead. $R^2_{\mathrm{bin}}$ is the in-sample binned conditional mean fit of Section~\ref{sec:methods}. The phase row is the fit under the causal trailing local baseline $\omega_0(t)$ of Eq.~\eqref{eq:beta_eff}; the amplitude row has no local baseline and is unaffected by that choice.}
\label{tab:sleep}
\end{table}

\begin{figure}[!ht]
\centering
\includegraphics[width=\textwidth]{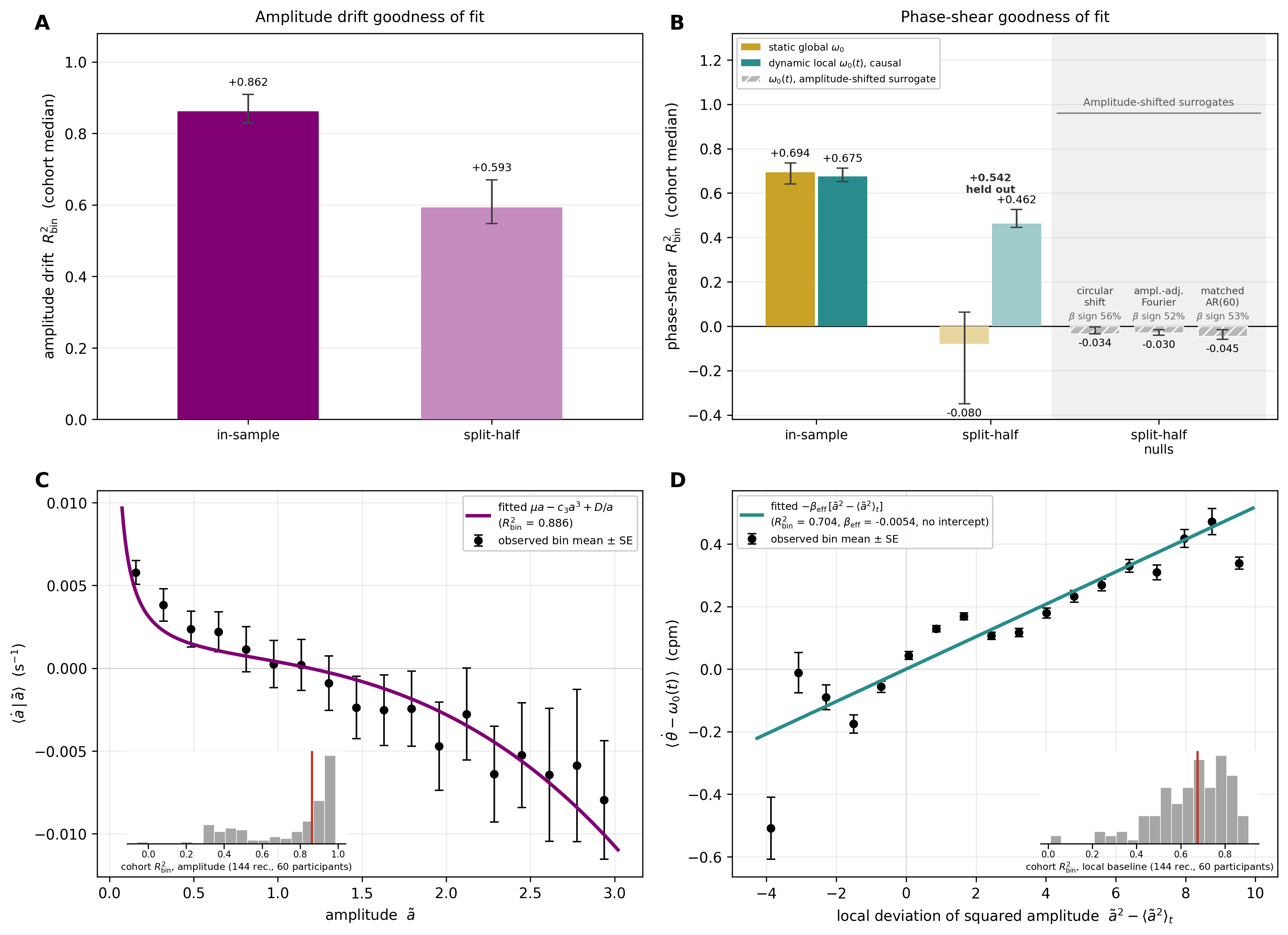}
\caption{\textbf{Conditional-drift agreement metric of the amplitude and phase-shear drift channels in the sleep cohort.} ({A}) Amplitude channel across the cohort: $R^2_{\mathrm{bin}}$ in sample and under split-half validation, as the median over the 60 participants. Most of the in-sample fit survives held out, 0.862 to 0.593. ({B}) Phase channel $R^2_{\mathrm{bin}}$s are plotted to compare the single global intercept $\omega_0$ (Eq.~\eqref{eq:pha}) with the causal trailing local baseline intercept $\omega_0(t)$ of Eq.~\eqref{eq:beta_eff}. For the split-half tests, the same local-baseline fit is compared to the three amplitude-shifted surrogates (shaded). The two intercepts are nearly indistinguishable in sample, and if anything favor the global one, 0.694 against 0.675, yet they separate by 0.542 held out, where the global intercept fails at $-0.080$. The surrogates place that held-out gain against its null: each preserves the autocorrelation, spectrum or marginal distribution of the real amplitude and destroys only its time alignment with $\dot\theta(t)$, and each collapses to $-0.034$, $-0.030$ and $-0.045$, respectively. Error bars in A and B are the 95\% bootstrap interval of the plotted cohort median, clustered on an individual participant; the two split directions are pooled to 288 held-out halves before the participant-level median is taken. ({C}) Amplitude drift of the representative recording of Fig.~\ref{fig:fig2}: increments $\Delta a/\Delta t$ binned by amplitude into 18 equal-width bins spanning the 0.5th-99.5th percentile of the state, shown as within-bin means $\pm$ standard error against the drift implied by that recording's own fitted $\mu$, $c_3$ and $D$ ($R^2_{\mathrm{bin}} = 0.886$). The rise at small $\tilde A$ is the $D/a$ term of Eq.~\eqref{eq:amp}. Inset: distribution of $R^2_{\mathrm{bin}}$ over the 144 recordings, with the participant-level cohort median 0.862 marked. ({D}) Phase drift of the same recording under the causal trailing local baseline, the fit the $\beta_{\mathrm{eff}}$ results use ($R^2_{\mathrm{bin}} = 0.704$, $\beta_{\mathrm{eff}} = -0.0054$; cohort median 0.675). }
\label{fig:fig3}
\end{figure}

\runin{Conditional-drift agreement metric.} The binned conditional-mean $R^2_{\mathrm{bin}}$ has cohort median 0.862 for the amplitude channel, with a median moving-block bootstrap 95\% interval of [0.744, 0.927] (Fig.~\ref{fig:fig3}A). For the phase channel, we first calculate the binned conditional-mean $R^2_{\mathrm{bin}}$ for the single global $\omega_0$ described in Eq. \ref{eq:pha}, which had a median of 0.685 ([0.388, 0.835]). Under split-half validation, the amplitude channel generalizes at a median $R^2_{\mathrm{bin}}$ of 0.593, while the phase under a single global $\omega_0$ does not, at $-0.080$.

\runin{Correcting for baseline-frequency drift.} As we have observed that the gastric-band frequency fluctuates (Fig. \ref{fig:fig2}D), the fitted $\omega_0$ is correct on average but wrong locally in a way that does not transfer to a held-out half of the recording. Replacing the global intercept with the slow causal local baseline of Eq.~\eqref{eq:beta_eff} raises the split-half $R^2_{\mathrm{bin}}$ from $-0.080$ to $+0.462$, a held-out gain of 0.542, positive in 90\% of held-out halves (Fig.~\ref{fig:fig3}B). The in sample $R^2_{\mathrm{bin}}$ of the two intercepts, scored on the same rows, are nearly indistinguishable and if anything favor the global one, 0.694 against 0.675, and thus, what separates them is generalization and not fit quality (Fig.~\ref{fig:fig3}B). The three amplitude surrogates collapse the effect while preserving everything else about the amplitude series. Against the dynamic local $\omega_0(t)$ split-half $R^2_{\mathrm{bin}}$ of $+0.462$, the circular time shift gives $-0.034$, the amplitude-adjusted Fourier surrogate $-0.030$, and the matched autoregressive series $-0.045$, with sign consistency at chance (56\%, 52\% and 53\%; Fig. \ref{fig:fig3}B). The fitted coupling is itself stable within a recording: $\beta_{\mathrm{eff}}$ carries the same sign in the early and the late half of the night in 98\% of recordings, and is negative in every participant.

\runin{Parameter estimates.} Cohort parameter estimates are reported in Table~\ref{tab:sleep}. The intrinsic rate is close to the canonical gastric value and tightly distributed, with $\omega_0$ median 3.14\,cpm (IQR [3.05, 3.24]), and it is the one parameter separated from zero by twice its own bootstrap standard error in every recording of every participant, at a median $|\hat\omega_0|/\mathrm{SE}$ of 104. The carrier is non-isochronous in a consistent direction: the raw shear is negative in 99\% of a participant's recordings on average and exceeds twice its own standard error in 96\% (median $\beta = -0.0049$), as does the local effective shear in sign (median $\beta_{\mathrm{eff}} = -0.0064$, negative in every recording). Under the stricter requirement that every night of a participant agree, which asks whether the sign is a property of the individual rather than of the night, the raw shear is negative on all nights of 97\% of participants and the effective shear on all nights of every participant. The amplitude-channel parameters separate from zero unevenly: the diffusion $D$ exceeds twice its standard error in 85\% of a participant's recordings and the cubic saturation $c_3$ in 41\%, but the growth rate $\mu$ sits at the origin (median $+1\times10^{-4}$) and exceeds twice its standard error only 2\%.

\subsection{Application to prandial cohort}
\begin{figure}[!ht]
\centering
\includegraphics[width=\textwidth]{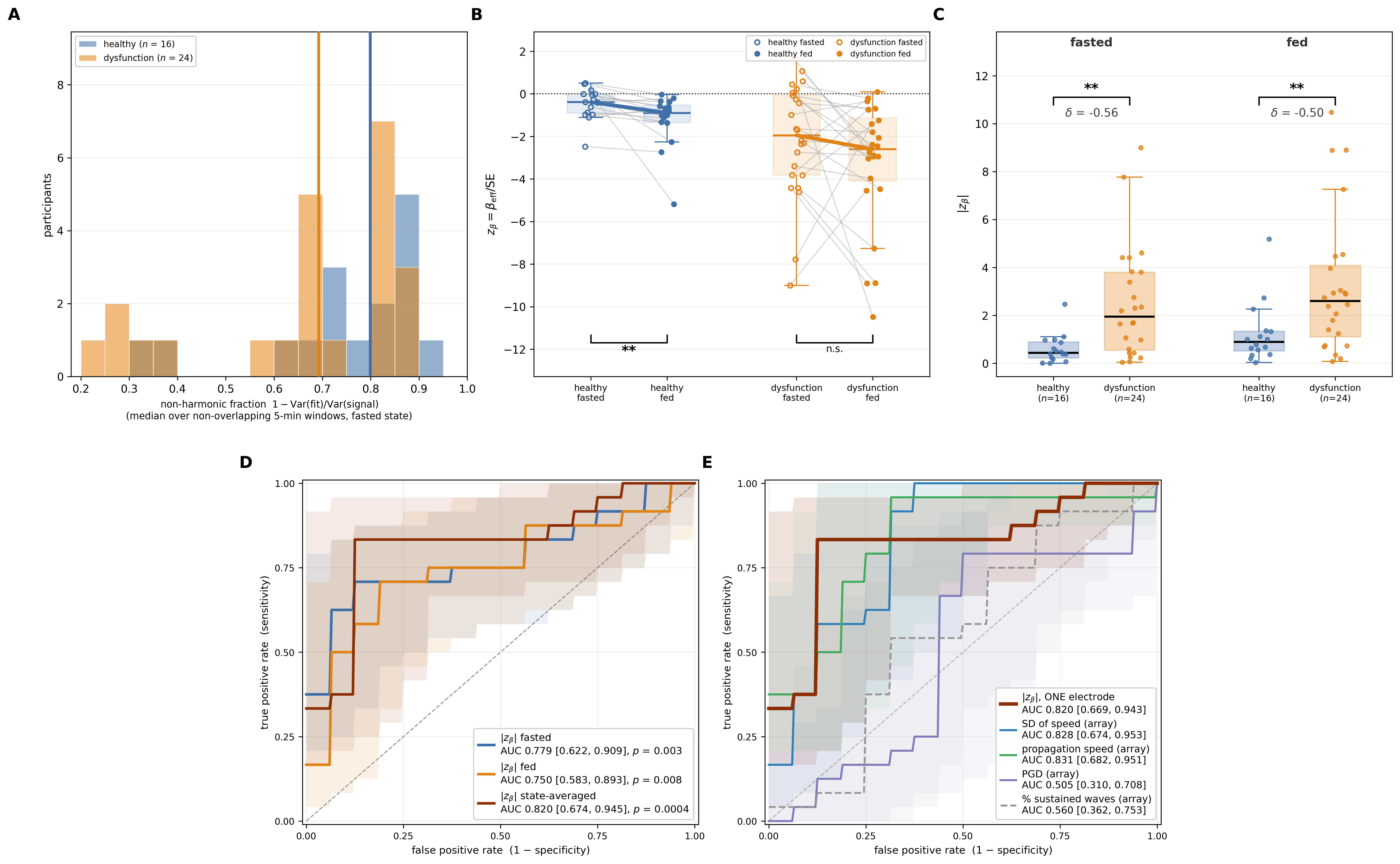}
\caption{\textbf{Parameter estimates of phase-shear in prandial cohort helps distinguish between healthy and unhealthy gastric dysfunction.} Healthy ($n = 16$ participants, 26 recordings) in blue, gastric dysfunction ($n = 24$ participants, one recording each) in orange throughout. Every plotted value is a participant's within-participant median, the repeated-measures unit of Section~\ref{sec:methods}.  Significance stars: $^{*}p<0.05$, $^{**}p<0.01$, $^{***}p<0.001$, n.s.\ otherwise. ({A}) Distribution of the non-harmonic fraction of gastric-band oscillator in the fasted state (similar to Fig. \ref{fig:fig2}A), shows healthy median at 0.80 [0.71, 0.86] versus dysfunction median 0.69 [0.52, 0.82] (Mann-Whitney $p = 0.066$, $\delta = 0.35$; the fed state gives $p = 0.0016$). ({B}) Paired fasted-to-fed $z_\beta$ for both groups. Brackets give the Wilcoxon signed-rank test. The healthy carrier deepens from a median $z_\beta$ of $-0.39$ to $-0.90$, in 13 of 16 participants (Wilcoxon $p = 0.0017$, sign test $p = 0.021$). Gastric dysfunction moves in the same direction, $-1.95$ to $-2.60$ in 18 of 24 participants, but does not reach significance by Wilcoxon ($p = 0.095$). ({C}) Between-group $|z_\beta|$ in each prandial state is presented. Boxes give median and IQR and brackets give the Mann Whitney test with Cliff's $\delta$. Fasted 0.48 versus 1.95 ($p = 0.0033$, $\delta = -0.56$); fed 0.90 versus 2.60 ($p = 0.0084$, $\delta = -0.50$). ({D}) Receiver operating characteristic curves distinguishing gastric dysfunction group against healthy, for each prandial state and for the within-participant average of the two states. Both curves in D and E present 95\% intervals from 10\,000 bootstrap resamples and a two-sided $p$ from the 10\,000 label permutations. ({E}) One electrode phase-shear measure compared against multi-electrode array features on the prandial cohort. Areas under the curve and their permutation $p$ values, with the Holm-Bonferroni correction across the seven array features, are given in Table~\ref{tab:array}.
}
\label{fig:fig4}
\end{figure}

The prandial cohort applies the model to shorter records under a controlled perturbation, and is where the parameters are tested to see if it can detect a change in gastric dynamics. The non-harmonic fraction for the prandial cohort has cohort median 0.80 ([0.71, 0.86]) in the 16 healthy participants in the fasted state, close to the 0.80 of the sleep cohort (Fig.~\ref{fig:fig4}A). The pooled dysfunction participants score lower, 0.69 ([0.52, 0.82]), suggesting that the gastric-band carrier signal is closer to a constant-amplitude isochronous sinusoid than healthy carriers are, although at the participant level the fasted contrast does not reach significance (Mann-Whitney $p = 0.066$, Cliff's $\delta = 0.35$) and only the fed state does ($p = 0.0016$, $\delta = 0.60$). We therefore treat the direction of this difference as exploratory; a lower-amplitude carrier is also more filter-dominated and so more nearly sinusoidal by construction, and a signal-to-noise-matched control is still owed.

\runin{Fit quality at this record length.} Parameter estimates for the full prandial cohort are reported in Table~\ref{tab:prandial}. The conditional-mean fit is weaker than in sleep, with amplitude $R^2_{\mathrm{bin}}$ median 0.29 fasted and 0.28 fed in the healthy participants, and 0.51 fasted and 0.63 fed in dysfunction. For the phase under the local baseline, the fit the $\beta_{\mathrm{eff}}$ results use, $R^2_{\mathrm{bin}}$ median in the healthy group is 0.19 fasted and 0.22 fed, and in dysfunction is 0.15 fasted and 0.31 fed. Record length rather than model form accounts for most of the difference: refitting the sleep data on short leading windows reproduces the same degradation within a single cohort and pipeline (amplitude $R^2_{\mathrm{bin}}$ 0.84 at 4\,hr, 0.70 at 60\,minutes, 0.57 at 30\,minutes and 0.35 at 10\,minutes), so that at the 30- and 60-minute block lengths of this cohort the sleep cohort itself scores 0.57 and 0.70 against the observed 0.42 and 0.56, closing roughly half to two thirds of the gap. The residual difference is small, and the ordering of fit quality across the three states (fasted 0.42, fed 0.56, sleep 0.86) tracks record length rather than diagnosis.
\begin{table}[!ht]
\centering
\footnotesize
\setlength{\tabcolsep}{3.5pt}

\resizebox{\textwidth}{!}{%
\begin{tabular}{@{}ll rrr rrr@{}}
\toprule
& & \multicolumn{3}{c}{Fasted} & \multicolumn{3}{c}{Fed} \\
\cmidrule(lr){3-5}\cmidrule(lr){6-8}
Parameter & & Median [IQR] & SE & Res. & Median [IQR] & SE & $|\hat\theta| > 2\,\mathrm{SE}$ \\
\midrule

\addlinespace
\multicolumn{8}{@{}l}{\textbf{Healthy ($n = 16$ participants, $n = 26$ recordings)}}\\
\multicolumn{8}{@{}l}{\textit{Phase channel}}\\
$\omega_0$ & intrinsic rate (cpm) & $3.177$ [$3.124$, $3.264$] & $0.093$ & 100\% & $3.248$ [$3.169$, $3.336$] & $0.063$ & 100\% \\
$\omega_0$ & intrinsic rate (rad\,s$^{-1}$) & $0.333$ [$0.327$, $0.342$] & $0.010$ & 100\% & $0.340$ [$0.332$, $0.349$] & $0.007$ & 100\% \\
$\beta$ & shear & $-2.9\times10^{-5}$ [$-5.5\times10^{-4}$, $1.0\times10^{-4}$] & $1.4\times10^{-3}$ & 10\% & $-1.9\times10^{-4}$ [$-1.1\times10^{-3}$, $-1.2\times10^{-5}$] & $3.9\times10^{-4}$ & 14\% \\
$z_\beta$ & standardized effective shear & $-0.392$ [$-0.900$, $-0.012$] & -- & 10\% & $-0.895$ [$-1.342$, $-0.523$] & -- & 19\% \\
\multicolumn{8}{@{}l}{\textit{Amplitude channel}}\\
$\mu$ & growth rate (s$^{-1}$) & $-0.0027$ [$-0.0034$, $-0.0022$] & $0.0055$ & 0\% & $-0.0018$ [$-0.0022$, $-0.0013$] & $0.0038$ & 0\% \\
$c_3$ & cubic saturation & $-3.5\times10^{-5}$ [$-2.9\times10^{-4}$, $8.0\times10^{-5}$] & $9.2\times10^{-4}$ & 0\% & $-4.9\times10^{-6}$ [$-3.6\times10^{-5}$, $6.4\times10^{-5}$] & $5.4\times10^{-4}$ & 0\% \\
$D$ & diffusion & $0.00203$ [$0.00164$, $0.00261$] & $0.00118$ & 25\% & $0.00174$ [$0.00130$, $0.00184$] & $0.00080$ & 52\% \\
\multicolumn{8}{@{}l}{\textit{Conditional-drift agreement metric}}\\
$R^2_{\mathrm{bin}}$ & amplitude & $0.295$ [$0.148$, $0.396$] & -- & -- & $0.279$ [$0.090$, $0.526$] & -- & -- \\
$R^2_{\mathrm{bin}}$ & phase, local $\omega_0(t)$ & $0.190$ [$0.011$, $0.341$] & -- & -- & $0.217$ [$0.141$, $0.387$] & -- & -- \\
\addlinespace
\multicolumn{8}{@{}l}{\textbf{Gastric dysfunction ($n = 24$ participants)}}\\
\multicolumn{8}{@{}l}{\textit{Phase channel}}\\
$\omega_0$ & intrinsic rate (cpm) & $3.160$ [$3.002$, $3.413$] & $0.105$ & 100\% & $3.217$ [$2.981$, $3.289$] & $0.079$ & 100\% \\
$\omega_0$ & intrinsic rate (rad\,s$^{-1}$) & $0.331$ [$0.314$, $0.357$] & $0.011$ & 100\% & $0.337$ [$0.312$, $0.344$] & $0.008$ & 100\% \\
$\beta$ & shear & $-7.2\times10^{-3}$ [$-1.1\times10^{-2}$, $4.1\times10^{-5}$] & $3.1\times10^{-3}$ & 46\% & $-7.5\times10^{-3}$ [$-1.4\times10^{-2}$, $-1.5\times10^{-3}$] & $3.4\times10^{-3}$ & 54\% \\
$z_\beta$ & standardized effective shear & $-1.947$ [$-3.812$, $-0.038$] & -- & 50\% & $-2.597$ [$-4.092$, $-1.115$] & -- & 62\% \\
\multicolumn{8}{@{}l}{\textit{Amplitude channel}}\\
$\mu$ & growth rate (s$^{-1}$) & $-0.0021$ [$-0.0030$, $-0.0009$] & $0.0057$ & 0\% & $-0.0015$ [$-0.0021$, $-0.0012$] & $0.0041$ & 0\% \\
$c_3$ & cubic saturation & $1.5\times10^{-4}$ [$1.5\times10^{-5}$, $3.9\times10^{-4}$] & $1.3\times10^{-3}$ & 0\% & $2.2\times10^{-4}$ [$1.1\times10^{-4}$, $3.6\times10^{-4}$] & $1.3\times10^{-3}$ & 0\% \\
$D$ & diffusion & $0.00187$ [$0.00127$, $0.00259$] & $0.00152$ & 12\% & $0.00149$ [$0.00123$, $0.00198$] & $0.00089$ & 33\% \\
\multicolumn{8}{@{}l}{\textit{Conditional-drift agreement metric}}\\
$R^2_{\mathrm{bin}}$ & amplitude & $0.507$ [$0.388$, $0.696$] & -- & -- & $0.632$ [$0.533$, $0.766$] & -- & -- \\
$R^2_{\mathrm{bin}}$ & phase, local $\omega_0(t)$ & $0.154$ [$0.001$, $0.390$] & -- & -- & $0.310$ [$0.202$, $0.480$] & -- & -- \\

\bottomrule
\end{tabular}%
}

\caption{\textbf{Stuart-Landau parameter estimates for the prandial cohort.} Stuart-Landau parameter estimates for the prandial cohort (40 participants: 16 healthy, 7 functional dyspepsia, 17 gastroparesis, from 50 recordings), reported as a repeated measure on the same convention as Table~\ref{tab:sleep}: each participant contributes the median of their own sessions, and the table gives the median [IQR] across participants. The patient arms are one recording per person, so their entries are unchanged by that step. The fasted and fed blocks of each recording are fitted separately, as described in Section~\ref{sec:methods}. The estimated parameters reported are the same as Table~\ref{tab:sleep}.}
\label{tab:prandial}
\end{table}

\runin{Use of standardized shear for prandial cohort.} Among the healthy participants alone, where no disease contrast exists, the raw index spreads 25-fold between system medians in the fasted state (Kruskal-Wallis $p = 0.082$ over 16 participants), whereas the standardized shear spreads 3.1-fold ($p = 0.28$); neither test is powered at this $n$, but the eightfold difference in spread is the relevant quantity, and it is a hardware effect by construction. Dividing each recording by its own block-bootstrap standard error removes the amplitude scale and weights each recording by how well its own shear is determined. We therefore report $z_\beta$ throughout this section, and all prandial results below are on that scale.

\runin{Effective shear deepens after a meal in healthy but not in gastric dysfunction.} In the 16 healthy participants, the standardized shear became more negative from the fasted to the fed block (Fig.~\ref{fig:fig4}B), with median $z_\beta$ moving from $-0.39$ to $-0.90$ and 13/16 participants shifting in that direction (paired Wilcoxon $p = 0.0017$, paired $t$ $p = 0.025$). Collapsing the repeated sessions sharpens rather than weakens this contrast, which is what is expected if the meal effect is a property of the individual that within-participant session noise was diluting. Within the gastric dysfunction group, one recording per person, the paired meal contrast is directionally consistent, with median $z_\beta$ moving from $-1.95$ to $-2.60$ and 18/24 participants shifting in that direction but not reaching significance (paired Wilcoxon $p = 0.095$; Fig.~\ref{fig:fig4}B).
\begin{table}[htbp]
\centering
\label{tab:array}
\begin{tabular}{lccc}
\hline
Measure & AUC [95\% CI] & $p$ & $p_{\mathrm{Holm}}$ \\
\hline
$|z_\beta|$, single electrode & 0.820 [0.669, 0.943] & 0.0006 & --- \\
\hline
Wave propagation speed & 0.831 [0.682, 0.951] & 0.0003 & 0.002 \\
SD of wave propagation speed & 0.828 [0.674, 0.953] & 0.0003 & 0.002 \\
Abnormal wave direction (\%) & 0.646 [0.469, 0.818] & 0.1227 & 0.613 \\
Inter-electrode phase-locking value & 0.615 [0.404, 0.815] & 0.2326 & 0.930 \\
Sustained waves (\%) & 0.560 [0.362, 0.753] & 0.5424 & 1.000 \\
Directional dispersion & 0.552 [0.367, 0.729] & 0.5922 & 1.000 \\
Phase-gradient directionality & 0.505 [0.310, 0.708] & 0.9632 & 1.000 \\
\hline
\end{tabular}
\caption{\textbf{Comparison of phase-shear measure against high-resolution multi-electrode array features.} Single-feature discrimination of gastric dysfunction ($n = 24$ participants) from healthy controls ($n = 16$ participants), for the single-electrode shear magnitude and for each multi-electrode array feature. The array measures are computed on a comparable central aperture across all three acquisition systems, so this is the same cohort as Fig.~\ref{fig:fig4}D and $|z_\beta|$ takes the same value in both. Areas under the curve carry a 95\% interval from 10\,000 stratified bootstrap resamples and a two-sided $p$ from 10\,000 label permutations. As the seven array features constitute one family of tests on a single cohort, $p_{\mathrm{Holm}}$ gives the Holm-Bonferroni-adjusted value across those seven. Only the two speed-derived array features survive correction.}
\end{table}
\runin{Magnitude of phase-shear distinguishes between healthy and gastric dysfunction recordings.} The magnitude of the standardized shear differs by roughly three- to fourfold: median $|z_\beta|$ 1.95 versus 0.48 fasted and 2.60 versus 0.90 fed (Mann-Whitney $p = 0.0033$ and 0.0084; Cliff's $\delta = -0.56$ and $-0.50$; Fig.~\ref{fig:fig4}C). We evaluated the standardized shear as a single-feature classifier of gastric dysfunction ($n = 24$ participants) against healthy controls ($n = 16$ participants). The magnitude of shear, $|z_\beta|$, gives an area under the receiver operating characteristic curve of 0.779 fasted (95\% CI [0.622, 0.909], $p = 0.0026$), 0.750 fed ([0.583, 0.893], $p = 0.0083$) and 0.820 ([0.674, 0.945], $p = 0.0004$) when the two states are averaged within participant (Fig.~\ref{fig:fig4}D). Leave-one-out cross-validation returns the same values to three decimal places for every marker above (0.820 for the state-averaged marker). These fits use the centered local baseline rather than the causal trailing one adopted for the overnight cross-validation, because a trailing window is poorly conditioned on records this short; refitting them causally degrades the in-sample phase fit substantially (healthy $R^2_{\mathrm{bin}}$ 0.23 to 0.04 fasted) but leaves the classification intact, at a state-averaged area under the curve of 0.836. By contrast, the fed-minus-fasted change in shear classifies at chance, $\Delta|z_\beta|$ giving 0.547 ($p = 0.63$). This analysis shows that the discriminating feature between healthy and unhealthy gastric function is therefore the level of the shear and not the shear during a meal response.

\runin{Comparison of phase-shear measure against high-resolution multi-electrode
array features.} To determine the further utility of the phase-shear measure extracted from one EGG electrode, we compared its performance to features extracted from multiple electrodes in high-resolution EGG, which primarily use phase-based measures \cite{gharibans2016high,hwangKuramotovonMisesTime2026,perleyGraphicalModelsEfficient2025}. High-resolution electrogastrography uses a cutaneous electrode array to estimate the direction and speed of the propagating slow wave, and spatial abnormalities in these measures have been reported to correlate with the severity of gastric symptoms
\cite{gharibansSpatialPatternsHighResolution2019c, ogradyAbnormalInitiationConduction2012a}. Seven array-derived measures were scored as single features against pooled gastric dysfunction, alongside $|z_\beta|$ on the same participants (Table~\ref{tab:array}, Fig.~\ref{fig:fig4}E). The two speed-derived features discriminate: wave propagation speed reaches an AUC of 0.831 (95\% CI [0.682, 0.951]) and the standard deviation of wave propagation speed 0.828 ([0.674, 0.953]), against 0.820 ([0.669, 0.943]) for $|z_\beta|$ from a single electrode. Neither differs from $|z_\beta|$ in a bootstrap of the AUC difference that resamples participants so that both measures see the same resample ($\Delta$AUC $-0.010$, [$-0.198$, $+0.180$] for wave propagation speed and $-0.010$, [$-0.148$, $+0.133$] for its standard deviation); the three measures lie within 0.011 of one another. The remaining five features do not separate the groups, with AUCs of 0.646 for \% abnormal wave direction, 0.615 for inter-electrode phase-locking value, 0.560 for \% sustained waves, 0.552 for directional dispersion and 0.505 for PGD, none of which survives correction for the seven comparisons. Therefore, a single-electrode measure of phase-shear, which incorporates dynamic information from both the amplitude and the frequency of the gastric oscillator, provides information about gastric function similar to that which a high-resolution multi-electrode array based on phase alone can provide.

\section{Discussion}
\label{sec:discussion}
This work introduces a statistical nonlinear dynamics framework for quantifying dynamic amplitude and frequency interrelationships in the waveforms from human noninvasive electrogastrography recordings. Rather than characterizing the gastric slow wave solely through static summaries such as dominant frequency or band power \cite{wolpertElectrogastrographyPsychophysiologicalResearch2020c}, we model the gastric-band analytic EGG as an effective stochastic oscillator and estimate a local, normalized measure of amplitude-dependent frequency modulation, $\beta_{\text{eff}}$. In long overnight recordings, this amplitude-frequency relationship generalized to held-out halves of each recording and was disrupted by surrogate amplitudes that preserved amplitude autocorrelation or spectral structure while breaking its temporal alignment with phase velocity. Applied at the participant level in the prandial cohort, effective shear varied with meal state and differed between healthy participants and the gastric dysfunction group. One possible interpretation is that elevated shear magnitude could reflect a
less stable relationship between gastric-band amplitude and local phase velocity, which could accompany the abnormal gastric  slow-wave initiation, conduction, and low-amplitude activity described in gastroparesis patients by high-resolution serosal gastric mapping \cite{ogradyAbnormalInitiationConduction2012a}. Furthermore, we demonstrate the single-electrode effective-shear measure showed discrimination between healthy and gastric-dysfunction participants that was comparable to, and in some analyses exceeded, that of multielectrode phase-based traveling-wave measures. Interestingly, multi-electrode array-derived speed variability and propagation speed were comparable to the phase-shear measure, and future work can try to quantify the effects amplitude dynamics have on propagating wave speed dynamics. Together, these findings demonstrate that amplitude-frequency coupling contains information about gastric-band dynamics not represented by features that solely use multi-variate phase or time-averaged spectral EGG metrics.

\runin{Limitations.} Our model is an effective description of processed, single-channel gastric-band EGG, not a direct measurement of a single gastric pacemaker or its underlying ICC network. Cutaneous EGG is a spatially aggregated observation of propagating activity and remains affected by electrode geometry, signal quality, motion, respiration, and preprocessing \cite{wolpertElectrogastrographyPsychophysiologicalResearch2020c, gharibansSpatialPatternsHighResolution2019c, ogradyOriginPropagationHuman2010} In particular, band-pass filtering is required to isolate the gastric band but systematically affects the scale and calibration of fitted quantities, including baseline frequency and raw shear. We therefore interpret $\beta_{\text{eff}}$ as a filter-conditioned relative dynamical phenotype rather than a directly identifiable biophysical coefficient. The prandial cohort also combines acquisition systems with different amplitude calibrations and includes a modest number of independent participants; the diagnostic and meal-related findings should therefore be regarded as preliminary associations requiring replication in prospectively acquired, same-device cohorts. Finally, the radial amplitude parameters $\mu$ and $c_3$ were weakly identified in filtered EGG data and were sensitive to record duration and the structured envelope dynamics of individual recordings. We consequently do not interpret them as direct measures of bifurcation proximity, self-sustainment, or physiological resilience \cite{strogatzNonlinearDynamicsChaos2018}. This limitation does not directly affect the effective-shear analysis, which is estimated from local covariation between phase velocity and median-normalized squared amplitude rather than from the fitted radial-drift parameters.

\runin{Future work.} These findings could be extended into multivariate analyses where one could model the spatial organization and propagation of the gastric slow wave across EGG channels rather than analyzing a selected single channel per recording. Such a framework could complement existing multichannel models of phase-based gastric dynamics by testing how amplitude dynamics shape local frequency and phase relationships \cite{perleyGraphicalModelsEfficient2025,hwangKuramotovonMisesTime2026}. A state-space formulation that explicitly represents latent oscillator states, structured nuisance processes, and measurement noise could jointly estimate the oscillator and observation processes, potentially reducing dependence on preprocessing choices and improving parameter identifiability \cite{StatisticalModelHuman}. Improved recovery of radial parameters, especially $\mu$ and $c_3$, and their relation to stochastic forcing, will be necessary before drawing conclusions about self-sustainment, bifurcation proximity, or resilience. If those parameters become reliably recoverable, phase-plane analyses may help characterize gastric dynamical trajectories and generate testable predictions about their response to physiological or meal-related perturbations. With these advances and external validation, amplitude-frequency coupling may become a useful dynamical biomarker of gastric state, meal response, and gastric dysfunction.

\section*{Acknowledgments}
This research was supported by the Wu Tsai Human Performance Alliance at Stanford University (to A.A.R. and T.P.C.), the Joe and Clara Tsai Foundation (to A.A.R. and T.P.C.), the Wu Tsai Neurosciences Institute (to T.P.C.), the Yu Faculty Scholar funds from the Stanford School of Engineering (to T.P.C.), and NIH grant 1U01DK140939 (to T.P.C.).

We thank Sarah F. Schoch, Martin Dresler and their team at the Donders Institute for Brain, Cognition and Behaviour (Radboudumc, Nijmegen, The Netherlands) for collecting, processing and contributing the overnight sleep physiology dataset. We additionally thank Leen Abdul Razzak and Hassler Bueno Garcia for collecting and contributing the prandial dataset.

\printbibliography

\end{document}